\RequirePackage{amsmath}
\documentclass[12pt,a4paper]{iopart}
\usepackage{graphicx}
\usepackage{color}
\usepackage{multirow}
\usepackage{epsfig}
\usepackage{epstopdf}
\usepackage{iopams}
\usepackage{cancel}
\usepackage{multicol}
\usepackage{ipa}
\usepackage{xcolor}
\usepackage{mathabx}

\usepackage[sorting=none, backend=biber, style=numeric-comp, sortcites=true, minbibnames = 10, maxbibnames = 11]{biblatex}

\usepackage[breaklinks=true,colorlinks=true,linkcolor=blue,urlcolor=blue,citecolor=blue]{hyperref}

\usepackage[deletedmarkup=xout]{changes}

\definechangesauthor[color=red]{RPB}
\definechangesauthor[color=cyan]{SW}
\definechangesauthor[color=orange]{MV}

\definecolor{olive}{RGB}{128,128,0}
\definechangesauthor[color=olive]{IK}
\definechangesauthor[color=purple]{JS}
\definechangesauthor[color=pink]{TB}

\begin{document}

\title[Electron dynamics of three distinct discharge modes]{Electron dynamics of three distinct discharge modes of a cross-field atmospheric pressure plasma jet}

\author{M. Klich$^{1}$, D. Schulenberg$^{2}$, S. Wilczek$^{3}$, M. Vass$^{2,4}$, T. Bolles$^{2}$, I. Korolov$^2$, J. Schulze$^{2}$, T. Mussenbrock$^{2}$ and R.P. Brinkmann$^{1}$}

\address{$^1$Chair for Theoretical Electrical Engineering, Faculty of Electrical Engineering and Information Technology, Ruhr University Bochum, D-44780,
Bochum, Germany}
\address{$^2$Chair of Applied Electrodynamics and Plasma Technology, Faculty of Electrical Engineering and Information Technology, Ruhr University Bochum, D-44780, Bochum, Germany}
\address{$^3$TH Georg Agricola University, D-44787, Bochum, Germany}
\address{$^4$ Institute for Solid State Physics and Optics, Wigner Research Centre for Physics, 1121 Budapest, Konkoly Thege Miklós str. 29-33, Hungary}

\date{\today}

\begin{abstract}
	This paper investigates the electron dynamics in three distinct discharge modes of a cross-field atmospheric pressure plasma jet, the COST-Jet.
    Thereby, the discharge modes are the non-neutral, the quasi-neutral, and the constricted mode.
    Using a hybrid Particle-In-Cell/Monte-Carlo Collisions (PIC/MCC) simulation, the study systematically varies the applied voltage and driving frequency to explore the operation modes and their relations.
    The results reveal that at low input power, the COST-Jet operates in a non-neutral mode, characterized by a discharge close to extinction, analogous to the chaotic mode observed in other plasma devices.
    As power increases, the jet transitions to a quasi-neutral mode, which aligns with the well-known $\Omega$- and Penning modes, comparable to the bullet mode in parallel-field jets.
    At the highest power levels, the COST-Jet enters a constricted mode, where the plasma significantly densifies and constricts towards the electrodes along the entire discharge channel.
    Experimental validation using phase-resolved optical emission spectroscopy (PROES) supports the simulation findings, particularly identifying the constricted mode as a distinct operational regime.
    These insights into the mode transitions of the COST-Jet under varying operational conditions help optimize plasma applications in various fields.
\end{abstract}

\begin{description}
    \item[Keywords:] plasma jet, operation modes, atmospheric pressure plasma, hybrid PIC/MCC simulation, electron dynamics, constricted plasma 
\end{description}

%%%%%%%%%%%%%%%%%%%%%%
%%%%%%%%%%%%%%%%%%%%%% Introduction
%%%%%%%%%%%%%%%%%%%%%%
\section{Introduction}

    Plasmas ignited under ambient conditions are among the first discharges ever investigated \cite{kogelschatz_dielectric-barrier_nodate, becker_non-equilibrium_2005}.
    Nevertheless, these discharges experience renewed interest and attention in recent times \cite{bruggeman_foundations_2017, kogelschatz_dielectric-barrier_nodate, tendero_atmospheric_2006, tachibana_current_2006}.
    Responsible for their re-establishment is the great flexibility in application.
    Technical non-equilibrium plasmas are generally known for their unique versatility \cite{lieberman_principles_2005, chabert_physics_nodate, becker_non-equilibrium_2005, raizer_gas_1997}.
    At atmospheric pressure, they offer solutions to pressing medical issues \cite{von_woedtke_foundations_2022}, urgent environmental questions \cite{bogaerts_co_2017, bogaerts_foundations_2022}, and many more \cite{bruggeman_foundations_2017, tachibana_current_2006, tendero_atmospheric_2006}.

    Regarding plasma creation, the number of plasma sources is almost as diverse as their applications \cite{bruggeman_foundations_2017, tendero_atmospheric_2006}.
    The numerous concepts of creating plasmas at atmospheric pressure further complicate the tedious task of comparing the results of different groups.
    Thus, their common standards and references are called for \cite{alves_foundations_2023}.
    One of these references is the European Cooperation in Science and Technology reference Microplasma Jet (COST-Jet) \cite{golda_concepts_2016, noauthor_mpns_2024}.
    The COST-Jet is a capacitively coupled, cross-field plasma jet (i.e., the gas stream is perpendicular to the direction of the electric fields).
    By design, this reference plasma source offers stability and comparability of the generated plasma \cite{golda_concepts_2016, riedel_reproducibility_2020}.
    Additionally, the rectangular geometry and the convenient optical access granted by two quartz windows ease both experimental and theoretical studies of the device.

    Driven by these advantages, the COST-Jet and similar plasma sources, like its predecessor the micro atmospheric pressure plasma jet ($\mu$APPJ) \cite{schulz-von_der_gathen_spatially_2008, hemke_spatially_2011}, have been the plasma source of choice for various studies.
    The electron dynamics are exceptionally well studied \cite{schroder_characteristics_2015, golda_comparison_2020, hemke_ionization_2012, bischoff_experimental_2018, schulenberg_mode_2024, korolov_control_2019, korolov_helium_2020, dunnbier_stability_2015, klich_simulation_2022} but also concepts to go beyond the status of a reference source have been looked into \cite{gorbanev_applications_2019}.
    Among the studies of the electron dynamics, several discharge modes of the COST-Jet are described and discussed.
    Many studies deal with the so-called $\Omega$- and Penning mode \cite{schulz-von_der_gathen_spatially_2008, bischoff_experimental_2018, gibson_disrupting_2019}.
    The $\Omega$-mode is comparable to the $\alpha$-mode of low-pressure plasmas.
    Here, electrons gain their energy by interacting with an expanding boundary sheath.
    The Penning mode is often compared to the $\gamma$-mode \cite{schulz-von_der_gathen_spatially_2008, schroder_characteristics_2015, golda_comparison_2020} due to similar locations and times of the main ionization peaks between the electrodes and within the RF period.
    Other studies \cite{dunnbier_stability_2015, zhang_excitation_2009, farouk_atmospheric_2008, laimer_glow_2006, laimer_investigation_2005, yang_comparison_2005, jianjun_shi_mode_2005}, however, describe that the $\gamma$-mode coincides with a constriction of the discharge towards the electrodes.
    By Golda et al. \cite{golda_comparison_2020}, this operation mode was labeled constricted mode.
    In more recent work, \cite{klich_simulation_2022}, we introduced the non-neutral regime as a distinct operation regime.

    Considering these findings on a general level, there are parallels to the works of Sun et al. \cite{sun_enhancement_2021} and Walsh et al. \cite{walsh_three_2010}.
    In their works, Sun et al. deal with a dielectric-barrier discharge (DBD), and Walsh et al. with a parallel-field atmospheric pressure plasma jet.
    The authors present three basic operation modes for their devices based on wide-ranged parameter studies.
    First, a low-power operation mode; second, an operation mode at intermediate conditions; and third, a high-power mode.
    Although the COST-Jet differs from these devices, we think the previously discussed operation modes of the COST-Jet can also be categorized into three distinct discharge modes.
    \begin{enumerate}
        \item[(I)] At low powers there is the non-neutral discharge mode \cite{klich_simulation_2022}.
            These discharges are comparable to the chaotic mode of Walsh et al. \cite{walsh_three_2010} in terms of being close to the extinction of the discharge.
        \item[(II)] With increasing input power, the COST-Jet reaches a quasi-neutral mode.
            Here, the distinction between $\Omega$- and Penning mode \cite{schroder_characteristics_2015, golda_comparison_2020, hemke_ionization_2012, bischoff_experimental_2018, schulenberg_mode_2024, korolov_control_2019, korolov_helium_2020, dunnbier_stability_2015} comes into play.
            This operation mode is comparable to the bullet mode of the parallel-field jet \cite{walsh_three_2010}.
        \item[(III)] At the upper end of the power scale, the COST-Jet can enter a constricted mode.
            As mentioned, the tendency of the discharge to constrict itself was reported previously \cite{dunnbier_stability_2015, zhang_excitation_2009, farouk_atmospheric_2008, laimer_glow_2006, laimer_investigation_2005, yang_comparison_2005, jianjun_shi_mode_2005, golda_comparison_2020}.
            However, these reports focus on just conceptionally identical plasma sources or find the constriction localized to the COST-Jet's nozzle.
            This work, aided by experimental data, shows that the constricted mode is a distinct operation mode that can ignite along the whole discharge channel.
            By the high-power nature of the constricted mode, it is comparable to the continuous mode found in the parallel-field jet \cite{walsh_three_2010}.
    \end{enumerate}

    In this work, we will apply a hybrid Particle-In-Cell/Monte-Carlo collisions (PIC/MCC) scheme \cite{eremin_new_2016, klich_simulation_2022} to simulate a helium/nitrogen mixture in the COST-Jet geometry for a wide range of parameters (i.e., by varying the applied voltage and the driving frequency).
    Doing so allows us to demonstrate three distinct operation modes for the cross-field driven plasma jet.
    Additionally, experimental data support the description and introduction of the constricted mode as the third distinct condition.
    The experimental data proves that a constricted plasma can exist in an operation mode that covers the entire discharge channel.

    The structure of the manuscript is as follows.
    The following section introduces the basis of the hybrid PIC/MCC simulation and presents details of the experimental setup that produced the data on the constricted mode.
    The results are discussed in section \ref{results}, which contains four subsections.
    Firstly, a general description of the discharge modes, their density profiles, and the connection between the modes is given in section \ref{general}.
    The following section \ref{exp_data} introduces the experimental evidence for the constricted mode we describe.
    Furthermore, a comparison between the simulated data and the experiment is made.
    Details of the electron dynamics during the transition from non-neutral discharges to quasi-neutral plasmas are compared in section \ref{Dynamics1}.
    In section \ref{Dynamics2}, a similar analysis is done for the electron dynamics of the constricted mode
    Section \ref{conclusion} summarizes the findings and presents some conclusions.

%%%%%%%%%%%%%%%%%%%%%%
%%%%%%%%%%%%%%%%%%%%%% Methods and Theory
%%%%%%%%%%%%%%%%%%%%%%
\section{Methods and models} \label{methods}

    \begin{figure}[t!]
        \centering
        \includegraphics[width=\textwidth]{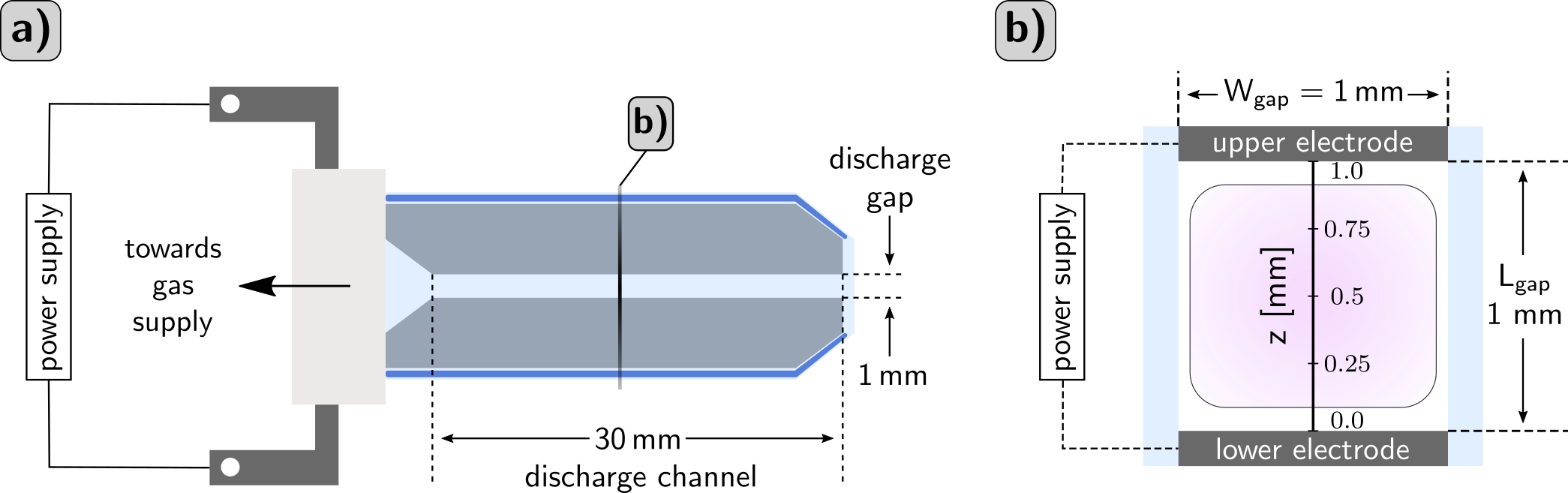}
        \caption{Simplified sketch of the COST-Jet (full characterization in \cite{golda_concepts_2016, noauthor_mpns_2024}).
            a) The cross-section sketched along the gas flow axis $x$,
            and b) the cross-section in a plane perpendicular to the gas flow.
            The black marking in panel a) shows the position of the cross-section sketched in panel b).
            The $z$-axis added in panel b) signifies the simulation axis of this work.}
        \label{fig:sketch}
    \end{figure}

    The plasma source modeled by the simulations conducted in this work is the COST-Jet \cite{golda_concepts_2016, noauthor_mpns_2024}.
    A detailed description of the source is found in previous work \cite{golda_concepts_2016, klich_simulation_2022, bischoff_experimental_2018, korolov_control_2019, korolov_helium_2020}.
    Figure \ref{fig:sketch} gives a sketch of the essential details.
    Additionally, the most relevant technical data is listed in the description of the experimental setup (see sec. \ref{experiment}).

    As in previous works \cite{klich_simulation_2022, bischoff_experimental_2018, korolov_control_2019, korolov_helium_2020, schulenberg_mode_2024}, one-dimensional simulations of the COST-Jet are conducted.
    Due to the relatively inert nature of the He/N\textsubscript{2}-mixture used in this work, a local (i.e. at one position along the gas flow axis) consideration of the plasma suffices to catch the relevant dynamics \cite{klich_simulation_2022, bischoff_experimental_2018, korolov_control_2019, korolov_helium_2020, schulenberg_mode_2024}.
    Thus, the simulations are conducted along the mid-axis of the jet's cross-section, referred to as the $z$-axis.
    The $z$-axis is marked in figure \ref{fig:sketch} panel b).

    The simulations are conducted with the simulation code \textit{Eehric} \cite{klich_simulation_2022}.
    \textit{Eehric} is an acronym for explicit electrostatic hybrid Particle-In-Cell.
    Electrons are treated by the kinetic Particle-In-Cell/Monte-Carlo collisions algorithm \cite{birdsall_particle--cell_1991, verboncoeur_particle_2005, donko_edupic_2021, wilczek_electron_2020, eremin_new_2016}.
    Ions are viewed as separate fluids, and the continuity equation is solved based on the drift-diffusion approximation for each ion species.
    The two parts of the simulation communicate via the Poisson equation and the species-specific source terms.
    A detailed description of the code was published in \cite{klich_simulation_2022}; the remainder of this manuscript will focus on the simulation parameters and relevant additions in the following section.

    Due to the underlying PIC/MCC scheme, the same constrictions in terms of grid size $\Delta z$ and timestep $\Delta t$ as for the PIC/MCC simulations apply \cite{turner_simulation_2013, wilczek_electron_2020, donko_edupic_2021, vass_revisiting_2022}.
    The spatial grid has to resolve the Debye length $\lambda_\mathrm{D}$ properly (i.e., $\lambda_\mathrm{D} \gtrsim 2 \Delta z$), thus for most cases, 201 grid points are sufficient.
    For the significantly elevated densities of constricted discharges, the number of grid points has to be drastically increased to 1500 for sufficient resolution of the Debye length. 
    Although constricted cases reach plasma densities in the order of $10^{19}\, \mathrm{m}^{-3}$, Coulomb collisions do not need to be considered.
    Hagelaar demonstrated that for an ionization degree $\alpha_\mathrm{i} < 10^{-5}$ Coulomb collisions do not significantly influence the particle transport \cite{hagelaar_coulomb_2016}.
    For the increased densities at hand, the atmospheric pressure counterbalances the plasma density in terms of the ionization degree (i.e., even for constricted plasmas $\alpha_\mathrm{i} \lesssim 10^{-6}$ is expected).

    There are two relevant criteria for the time step $\Delta t$.
    First of all, the time step has to be sufficiently small to resolve oscillations on the scale of the electron plasma frequency $\Delta t\, \omega_\mathrm{e} \lesssim 0.2$ \cite{turner_simulation_2013, wilczek_electron_2020, donko_edupic_2021, vass_revisiting_2022}.
    For low and intermediate-pressure applications, this criterion dictates the maximal time step.
    However, collisions become the limiting factor at high pressure, and the plasma frequency is usually automatically resolved when following the second criterion.
    This second criterion arises from the necessity of the PIC/MCC algorithm to avoid consecutive collisions at the same timestep.
    Each simulated superparticle cannot collide more than once per time step $\Delta t$, which results in a timestep of $\Delta t \approx 4.79 \times 10^{-14}\,$s.

    Apart from the varied parameters (i.e., driving voltage or frequency), all simulated cases share the same primary conditions.
    The gas pressure $p$ is fixed to $101325\,$Pa, and gas heating is neglected.
    Thus, the gas temperature has a constant value $T_\mathrm{g} = 300\,$K.
    (Please note that the neglect of gas heating is a crude assumption.
    However, the gas temperature is treated as an input parameter to the simulation.
    As long as the gas temperature is not self-consistently calculated during the simulation, it is another parameter that, as previous work showed \cite{schulenberg_mode_2024}, influences mode transitions.)
    The background gas is a mixture of helium and nitrogen with a fixed nitrogen admixture $x_\mathrm{N_2} = 0.001$.
    The electrodes are considered to be perfectly absorbing for any incoming electrons.
    Neumann boundary conditions are assumed for ions, and ion-induced secondary electron emission is considered by constant coefficients $\gamma_s$.
    The values are $\gamma_\mathrm{He^+} = 0.25$, $\gamma_\mathrm{He_2^{\ +}} = 0.23$, $\gamma_\mathrm{N_2^{\ +}} = 0.11$, and $\gamma_\mathrm{N_4^{\ +}} = 0.1$.
    These values are based on an empirical formula (details given in previous work \cite{raizer_gas_1997, klich_simulation_2022}).

%%%%%%%%%%%%%%%%%%%%%%
%%%%%%%%%%%%%%%%%%%%%% Chemistry
%%%%%%%%%%%%%%%%%%%%%%
\subsection{Discharge chemistry} \label{chemistry}
    
   The chemistry set used in this work is an extended version of the set used in our previous work \cite{klich_simulation_2022}.
   Among the modifications, the three most notable are the change of the cross-section set for electrons colliding with nitrogen molecules, the addition of the N\textsubscript{4}$^+$-cluster ion, and the implementation of chemical loss channels for all species.
   A detailed record of the changes to the plasma chemistry compared to \cite{klich_simulation_2022} is given in table \ref{tab:chemistry}.
   Added and modified reactions are marked with a star behind the number (e.g., $29^\ast$).
   The non-modified first 28 reactions are skipped.

   This work describes the collisions via electrons and nitrogen molecules by cross-sections retrieved from the SIGLO database \cite{noauthor_siglo_2024} via the LXCat website.
   These cross sections are based on measurements by Phelps and Pitchford \cite{phelps_anisotropic_1985} and treat the ionization via a single effective state (see 29$^\ast$ in tab. \ref{tab:chemistry}).
   Since separating different excited states of the nitrogen ions is not necessary for our work, using the SIGLO data proved the most efficient in terms of computational load.

   \begin{table}[t!]
		\centering
		\begin{tabular*}{1.0\textwidth}{ l @{\extracolsep{\fill}} l l c c}
			\hline
			No.~ & reaction & process name & $\varepsilon_\mathrm{thr}$ / $k_\mathrm{r}$ &  src. \\
			\hline
            $\vdots$ & $\vdots$ & $\vdots$ & $\vdots$ & $\vdots$\\
            29$^\ast$ & $\mathrm{e} + \mathrm{N}_2 \rightarrow \mathrm{e} + \mathrm{N\textsubscript{2}\textsuperscript{+}} + \mathrm{e}$ & ionization & $15.6\, \mathrm{eV}$ & (1)\\
            30$^\ast$ & $\mathrm{e} + \mathrm{N_4}^+ \rightarrow \mathrm{N_2} + \mathrm{N_2}$ & recombination & - & (2) \\ %$2.0 \times 10^{-12}\, \sqrt{\frac{T_0}{T_\mathrm{e}}}\ \mathrm{\frac{m^3}{s}}^\ast$ & (3)\\
			\hline
			31 & $\mathrm{He}^\ast + \mathrm{N}_2 \rightarrow \mathrm{He} + \mathrm{N\textsubscript{2}\textsuperscript{+}} + \mathrm{e}$ & Penning ionization & $5.0 \times 10^{-17}\, \mathrm{\frac{m^3}{s}}$ & \cite{sakiyama_corona-glow_2006, martens_dominant_2008}\\
			32 & $\mathrm{He}^+ + \mathrm{He} + \mathrm{He} \rightarrow \mathrm{He\textsubscript{2}\textsuperscript{+}} + \mathrm{He}$ & ion conversion & $1.1 \times 10^{-43}\, \mathrm{\frac{m^6}{s}}$ & \cite{brok_numerical_2005, sakiyama_corona-glow_2006, martens_dominant_2008}\\
            33$^\ast$ & $\mathrm{He_2}^+ + \mathrm{N_2} \rightarrow \mathrm{He_2}^\ast + \mathrm{N_2}^+$ & charge exchange & $1.4 \times 10^{-15}\, \mathrm{\frac{m^3}{s}}$ & \cite{brok_numerical_2005, martens_dominant_2008}\\
            34$^\ast$ & $\mathrm{N_2}^+ + \mathrm{N_2} + \mathrm{He} \rightarrow \mathrm{N_4}^+ + \mathrm{He}$ & ion conversion & $1.9 \times 10^{-41}\, \mathrm{\frac{m^6}{s}}$ & \cite{martens_dominant_2008, gaens_kinetic_2013}\\
            35$^\ast$ & $\mathrm{N_4}^+ + \mathrm{He} \rightarrow \mathrm{N_2}^+ + \mathrm{N_2} + \mathrm{He}$ & dissociation & $2.5 \times 10^{-21}\, \mathrm{\frac{m^3}{s}}$ & \cite{martens_dominant_2008, gaens_kinetic_2013} \\
            \hline
            36$^\ast$ & $\mathrm{e} + \mathrm{He} \rightarrow \mathrm{e} + \mathrm{He \left(^3S_1 \right)}$ & el. excitation & $22.92\,$eV & (3)
		\end{tabular*}
		\caption{Plasma chemical reactions considered in the simulation.
            The last column cites the data sources and uses the following abbreviations if no direct citation is given: 
            (1) data based on \cite{phelps_anisotropic_1985} obtained from the SIGLO database \cite{noauthor_siglo_2024} on LXCat,
            (2) a cross-section estimated based on the rate coefficient given by Kossyi et al. \cite{kossyi_kinetic_1992}. Details on the estimation are provided in \cite{kuhfeld_picmcc_2023},
            (3) data based on \cite{bartschat_electron-impact_1998, ralchenko_electron-impact_2008} obtained from the BIAGI database \cite{biagi_biagi_2024} on LXCat.
            The second column from the right gives threshold energies $\varepsilon_\mathrm{thr}$ for electron reactions and reaction rates $k_\mathrm{r}$ for ion reactions.}
		\label{tab:chemistry}
	\end{table}

    % In our previous work \cite{klich_simulation_2022}, the role of N\textsubscript{4}$^+$-ions was neglected.
    The N\textsubscript{4}$^+$-cluster-ion is considered and generated by the ion-conversion of N\textsubscript{2}$^+$ (s. 34 in tab. \ref{tab:chemistry}).
    For the $0.1\%$ nitrogen used in this work, the N\textsubscript{4}$^+$-ion plays a major to a dominant role as predicted by the study of Martens et al. \cite{martens_dominant_2008}.
    The low nitrogen mixture justifies the use of the values of the ion mobility of N\textsubscript{4}$^+$ in pure He as a good approximation for the mobility $\mu_\mathrm{i,N\textsubscript{4}^+}$.
    The necessary data is obtained from the mobility data collection of Viehland and Mason \cite{viehland_transport_1995}.

    As for the loss processes, it proved necessary to include chemical loss channels for all species to simulate the high-power cases of the constricted mode.
    Based on their reaction rates, processes 33-35 were added as source/loss terms in the ion continuity equations.
    For the electron-N\textsubscript{4}\textsuperscript{+}-ion-recombination (s. 30$^\ast$), an estimation for the cross section was done based on the rate coefficient given by Kossyi et al. \cite{kossyi_kinetic_1992}.
    The same estimation is used by Kuhfeld et al. \cite{kuhfeld_picmcc_2023} and results in an electron impact cross section $\sigma_\mathrm{e/N_4^{\ +}} [\mathrm{m^2}] = 4.81 \times 10^{-19} / \epsilon$.
    In this formula, $\epsilon$ denotes the electron energy in eV.
    The estimated cross-section is then used in the electron null-collision scheme \cite{skullerud_stochastic_1968, nanbu_probability_2000, klich_simulation_2022}.

    For the comparison with the experimental data (see next section), the excitation to the He$\left( ^3S_1 \right)$ is included (process 36, tab. \ref{tab:chemistry}).
    The cross-section data is retrieved from the Biagi database \cite{biagi_biagi_2024} and based on the data of Bartschart \cite{bartschat_electron-impact_1998} and Ralchenko et al. \cite{ralchenko_electron-impact_2008}.

%%%%%%%%%%%%%%%%%%%%%%
%%%%%%%%%%%%%%%%%%%%%% Experimental stuff
%%%%%%%%%%%%%%%%%%%%%%
\subsection{Experimental setup} \label{experiment}

    In this work, experiments are performed using an RF-driven reference microplasma COST-jet \cite{golda_concepts_2016, noauthor_mpns_2024} operated in helium (5.0 purity) with a fixed 0.1\% admixture of nitrogen (5.0 purity).
    The flow rates are fixed at $1000\,$sccm for helium, and 1$\,$sccm for nitrogen. 
    The gas handling system has been previously described in \cite{korolov_control_2019, schulenberg_mode_2024}.
    The jet consists of two identical stainless steel electrodes covered by two quartz plates, confining the plasma volume to $1\times1\times30$ mm$^3$.
    The jet is operated at atmospheric pressure with a single driving frequency of $13.56\,$MHz, for which an internal coupling circuit is utilized for impedance matching.
    The voltage signal is generated by a waveform generator (Keysight 33600A), amplified by a broadband power amplifier (Vectawave VBA250-400), and applied to the powered electrode through the matching network.
    The other electrode is grounded. The voltage signal at the powered electrode is measured and monitored using a high-voltage probe (Tektronix P6015A, 75 MHz) and a USB oscilloscope (Picoscope 6402C, 250 MHz, 5 Gs/s). 

    The phase-resolved optical emission spectroscopy (PROES) technique is used to validate the simulation findings in this work.
    The PROES measurements are conducted using an ICCD camera (4 Picos from Stanford Computer Optics) equipped with a telecentric lens (Edmund Optics, 86-658).
    The ICCD camera is synchronized with the arbitrary waveform generator.
    The lens is equipped with a narrowband interference filter (Thorlabs, 700 nm, FWHM: 15 nm) to monitor the He emission line at 706.5 nm.
    The central position of the images is located in the middle of the constricted active plasma region.
    To obtain spatio-temporal plots, the images are binned in the direction perpendicular to the electrode gap.
    The temporally and spatially-resolved electron-impact excitation rate from the ground state into the He-I (3s) $^3$S$_1$ state is calculated from the measured intensity using a collisional-radiative model \cite{schulze_phase_2010}, assuming an effective lifetime of 5 ns \cite{west_optical_nodate}.
    The energy threshold for this electron impact excitation process is 22.7 eV, meaning that energetic electrons above this energy are traced.
    Contributions from other lines or bands within the filter's transmission range, as well as from cascade processes from higher-lying levels, are considered negligible.

    The main purpose of the experimental measurement in this work is to validate simulation findings regarding the origin of the constricted mode by comparing simulation results of the excitation rates with experimental data.
    Due to the extensive heat production in such a discharge mode, performing stable measurements in the original COST-jet configuration is challenging.
    Therefore, electrode modification is required to achieve a stable enough discharge to perform the PROES measurements. For a detailed discussion, see section \ref{exp_data}.
    
%%%%%%%%%%%%%%%%%%%%%%
%%%%%%%%%%%%%%%%%%%%%% Results
%%%%%%%%%%%%%%%%%%%%%%
\section{Results and Discussion} \label{results}

%%%%%
\subsection{General description} \label{general}
	\begin{figure}[t!]
        \centering
		\includegraphics[width=\textwidth]{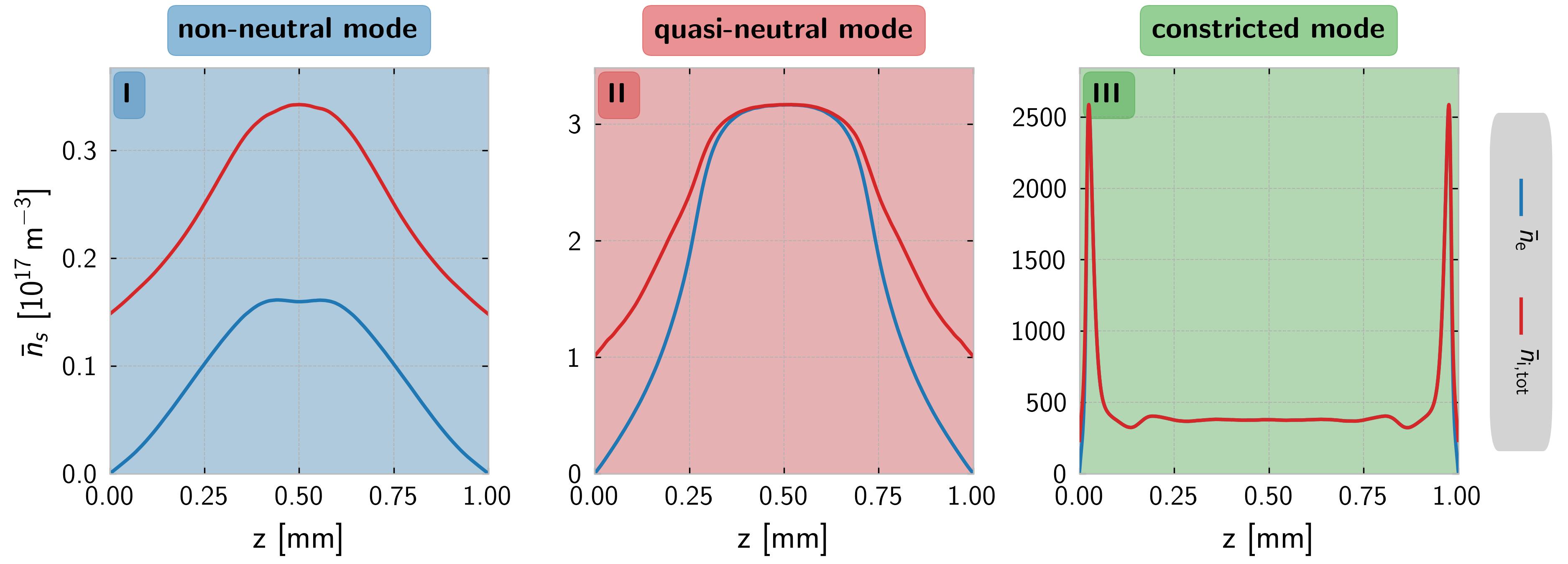}
		\caption{Time averaged and spatially resolved density profiles of electrons (blue) and ions (red) in the three operation modes.
            a) non-neutral mode (I) at $f_\mathrm{RF} = 12\,$MHz and $V_\mathrm{RF} = 200\,$V,
            b) quasi-neutral mode (II) at $f_\mathrm{RF} = 28\,$MHz and $V_\mathrm{RF} = 220\,$V, 
            and c) constricted mode (III) at $f_\mathrm{RF} = 34\,$MHz and $V_\mathrm{RF} = 400\,$V.
            All cases share the same base parameters: $p = 101325\,$Pa, $x_\mathrm{N_2} = 0.001$, $T_\mathrm{g} = 300\,$K.\\
            (Animated versions of similar profiles are provided as additional material.)} 
		\label{fig:ave_densities}
    \end{figure}
 
    The COST-Jet generally offers three distinct operation modes.
    These discharge modes vastly differ in their plasma densities, absorbed powers, and discharge dynamics.
    This section provides an overview of the operation modes before profoundly delving into the dynamics' details.
    Apart from presenting general information, the question of how to operate a chosen mode will be tackled.
    
    Figure \ref{fig:ave_densities} offers an overview of the temporally averaged spatial density profiles of electrons (blue) and ions (red) in these discharge regimes.
    Panel I shows the non-neutral mode (regime I), II the quasi-neutral mode (regime II), and panel III presents the constricted mode (regime III).
    As visible by the vastly different operation conditions (i.e., I: $f_\mathrm{RF} = 12\,$MHz $V_\mathrm{RF} = 200\,$V, II: $f_\mathrm{RF} = 28\,$MHz $V_\mathrm{RF} = 220\,$V, and III: $f_\mathrm{RF} = 34\,$MHz $V_\mathrm{RF}=400\,$V), there is a non-trivial answer to the previously raised question.
    We will first focus on a general description of the operation modes and then discuss their relation later.

    At first glance at the ordinates, it is visible that figure \ref{fig:ave_densities} covers five orders of magnitude in terms of plasma density.
    At low densities, the discharge is struggling to exist.
    This struggle defines the characteristics of the non-neutral mode.
    Electrons are highly mobile and organized within a group of approximately Gaussian shape (by the temporal integration, this Gaussian turns into the visible bimodal profile - c.f., I).
    As discussed in \cite{klich_simulation_2022}, the Gaussian electron density profile oscillates back and forth between the electrodes within each RF period.
    Quasi-neutrality can be fulfilled locally at specific times within the RF period, but on time-average electron and ion densities deviate.

    Panel II of figure \ref{fig:ave_densities} shows the classical density profile of capacitively coupled discharges.
    In time average, there is a quasi-neutral bulk region in the center of the discharge surrounded by the two boundary sheath regions.
    The existence of a distinct bulk region is characteristic of the quasi-neutral mode.
    Due to the small dimensions of the COST-Jet and the high pressure, high plasma densities in the order of $10^{17}\,$m\textsuperscript{-3} are necessary for a sheath-bulk-structure to be established.
    Thus, the quasi-neutral mode can solely exist for sufficiently high plasma densities.
    The transition between the previous non-neutral discharges and the quasi-neutral plasma is smooth.
    Electron and ion densities gradually grow closer, a tiny and highly transient bulk region appears, and eventually, the bulk-sheath structure stabilizes. 
    Later on, this transition will be referred to as regime I/II and its dynamics compared to the non-neutral and quasi-neutral modes.
    
    The quasi-neutral discharges of panel II and those close to quasi-neutrality are best investigated.
    The modes of power absorption, namely the $\Omega$- and the Penning-mode \cite{hemke_ionization_2012, schroder_characteristics_2015, schulz-von_der_gathen_spatially_2008, dunnbier_stability_2015}, that are unique to high-pressure discharges are investigated in these plasmas.
    Especially the transition to the $\gamma$-mode-like Penning-mode \cite{bischoff_experimental_2018, schulenberg_mode_2024,dunnbier_stability_2015}, can only occur when there are distinct plasma bulk and boundary sheaths.
    The power dynamics will be discussed in section \ref{Dynamics1}.

    Panel III shows the density profile of a constricted discharge.
    As reported before \cite{zhang_excitation_2009}, high-pressure discharges tend to constrict themselves to minuscule regions before the electrodes.
    In terms of plasma density, this characteristic coincides with breaking the usual diffusion profile of the densities.
    Constricted plasmas establish local electron and ion density maxima in close proximity to the electrodes.
    Combined with the extremely high plasma density, these discharges are characterized by tiny sheath regions that shield a large and steady quasi-neutral plasma bulk.
    Similar density profiles have been reported by Farouk et al. \cite{farouk_atmospheric_2008}.
    
    Transitioning between the operation modes is closely related to influencing the plasma density by any means.
    For this work, the transition between the discharge regimes will be triggered by either raising the driving frequency or the driving voltage.
    Both are known to influence the plasma density, and the corresponding relations will be discussed in the following \cite{lieberman_principles_2005}.
    However, constricted discharges require a critical field strength in front of the electrodes.
    Comparable to the ignition voltage of a direct-current (dc) discharge \cite{lieberman_principles_2005, raizer_gas_1997, gudmundsson_foundations_2017}, secondary electrons drive the constricted mode.
    Thus, the critical field strength is reached when secondary electrons cause sufficient ionization inside the sheath.
	
	\begin{figure}[t!]
		\centering
		\includegraphics[width=\textwidth]{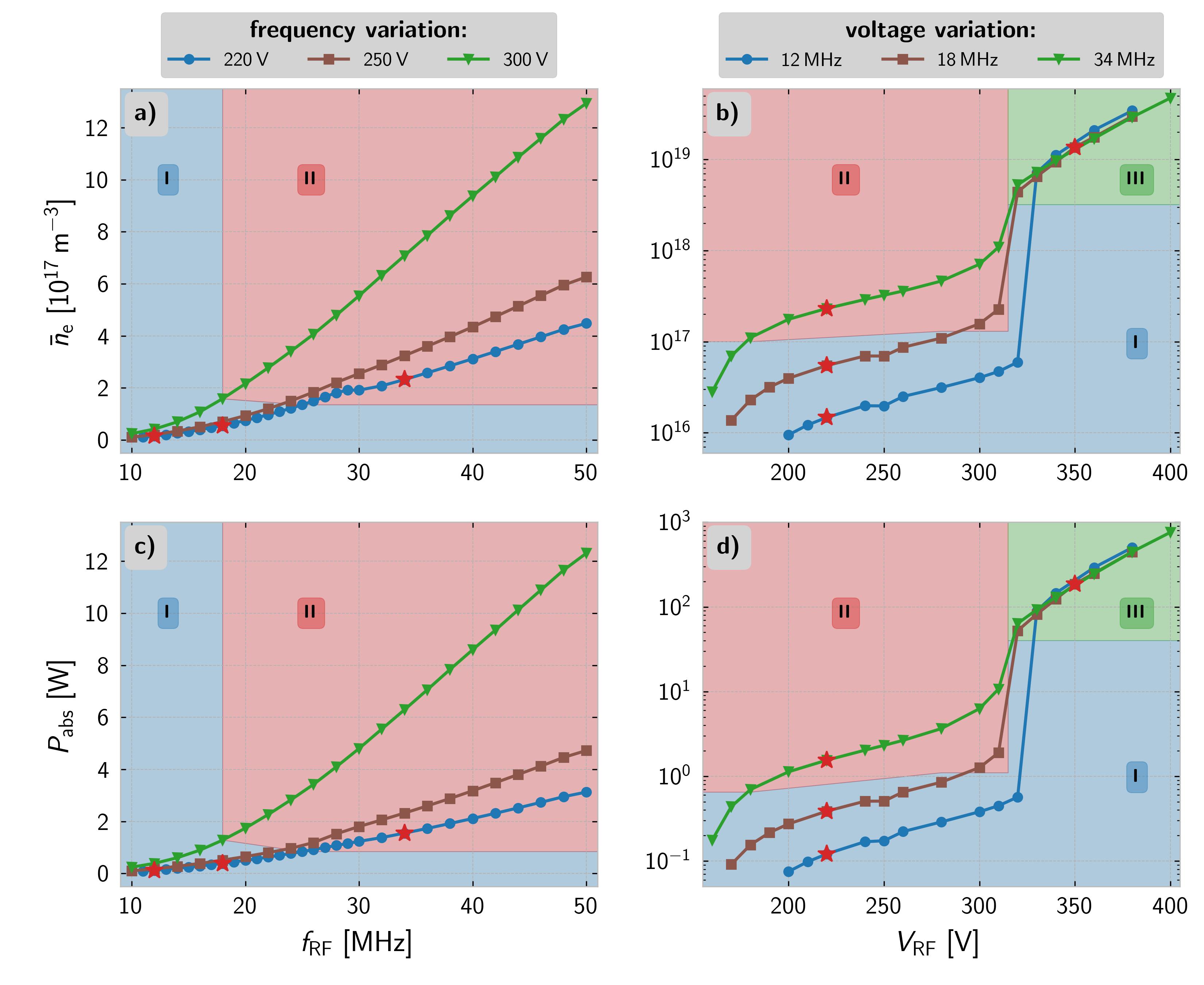}
		\caption{Temporally and spatially averaged plasma density $n_\mathrm{e}$ and absorbed power $P_\mathrm{abs}$ as a function of driving frequency $f_\mathrm{RF}$ (panels a) and c)) and voltage amplitude $V_\mathrm{RF}$ (panels b) and d)).
            The colored polygonic shapes in the background approximately mark where each panel's operation modes are located.
            I: non-neutral mode (blue), II: quasi-neutral mode (red), and III: constricted mode (green).
            The red stars mark the cases where sections \ref{Dynamics1} and \ref{Dynamics2} will provide a deeper analysis.
            All cases share the same base parameters: $p = 101325\,$Pa, $x_\mathrm{N_2} = 0.001$, $T_\mathrm{g} = 300\,$K.}
		\label{fig:parameters}
	\end{figure}

    Both voltage and frequency variations are used to determine how the operation modes correlate.
    Figure \ref{fig:parameters} presents the results of these variations.
    It shows both the spatially and temporally averaged electron density $\bar{n}_\mathrm{e}$ (top row) and the absorbed power $P_\mathrm{abs}$ (bottom row).
    The left column presents both as a function of the driving frequency $f_\mathrm{RF}$, and the right column gives them in dependence of the driving voltage $V_\mathrm{RF}$.
    The absorbed power $P_\mathrm{abs}$ is estimated by calculating the spatially and temporally averaged total power density $\langle p_\mathrm{abs} \rangle = \langle \vec{j}_\mathrm{tot} \cdot \vec{E} \rangle$, and multiplying the resulting value with $30\, \mathrm{mm^3}$ (the volume of the COST-Jet's discharge channel, cf. sec. \ref{experiment}).
    Both the frequency and the voltage variation were done for three different values of the respective other parameter.
    For the frequency variation (panels a) and c)), the variation has been done for three different voltages (blue $220\,$V, brown $250\,$V, and green $300\,$V).
    For the voltage variation (panels b) and d)), three frequencies (blue $12\,$MHz, brown $18\,$MHz, and green $34\,$MHz) have been investigated.

    Additionally, the colored background serves as a reference.
    Each operation mode is represented by one color (I: non-neutral blue, II: quasi-neutral red, and III: constricted green).
    These markings are extrapolations based on the simulated data.
    The criterion for a case to lie inside the blue canvas is $\bar{s} \gtrsim 0.5\,$mm, where $\bar{s}$ is the mean sheath width.
    The sheath width's introduction and significance will be discussed later.
    The steep increase in the density curves identifies the constricted mode.
    For a density increase $\frac{\Delta n_\mathrm{e}}{\Delta V} \gtrsim 4.1 \times 10^{17}\, \mathrm{\frac{m^{-3}}{V}}$, we consider all following cases to be constricted.
    (Searching for the characteristic bimodal time-averaged density profile in the simulation data helps to verify the transition.)
    The most important information visible by these marks is that neither panel a) nor c) has any green canvas.
    As mentioned before, the constricted mode requires a critical voltage.
    This voltage is comparable to the breakdown voltage of a dc discharge \cite{lieberman_principles_2005, raizer_gas_1997} (more on the analogies between the constricted mode and a dc discharge later).
    By extrapolating the data from panel a), one finds that raising the frequency can produce plasma densities similar to those shown for the constricted mode in panel b) and, similarly, the absorbed power (cf. panels c) and d)).
    Consequently, one might expect to encounter constricted discharges at higher frequencies ($f_\mathrm{RF} > 50\,$MHz), too.
    However, the simulated cases at $f_\mathrm{RF} = 50\,$MHz do not show any signs of being close to the transition to constriction (the voltage variation exhibits local density maxima inside the sheath when the discharge is close to the II-III transition).
    We understand these results by the varied time scales.
    Although the plasma densities are comparable and electrons experience similar field strength provoked by the according space charges, these fields influence the electrons for decreasingly smaller periods when the frequency is raised.
    Thus, the enormous space charge fields cause ionization cascades inside the sheath.
    Yet, the sheath collapses too early for these ionization cascades to build up the critical densities in front of the electrode that would cause constriction of the discharge.
    We additionally ran preliminary simulations at $f_\mathrm{RF} = 80$ and $100\,$MHz that confirmed this interpretation.
    However, we decided to limit the data shown in this work to $f_\mathrm{RF} \lesssim 50\,$MHz to ensure that electromagnetic effects are negligible \cite{lieberman_standing_2002, eremin_modeling_2023}.

    Overall, the comparison between the upper and lower row of figure \ref{fig:parameters} shows that $\bar{n}_\mathrm{e}$ and $P_\mathrm{abs}$ exhibit the same scaling with frequency and voltage.
    These scalings are known from literature \cite{lieberman_principles_2005} and present as follows.
    For density and absorbed power as a function of frequency, panels a) and c) of figure \ref{fig:parameters} show a quadratic trend ($\bar{n}_\mathrm{e} \propto f_\mathrm{RF}^2$).
    This is the expected scaling.
    Yet, the scaling proves useless in determining the operation mode and coloring the canvas.
    Nevertheless, varying the frequency influences the density, and both the non-neutral and the quasi-neutral modes occur.
    The transition between them is smooth.
    Defining characteristics of the non-neutral discharge fade away while the plasma bulk slowly establishes.

    There are three different slopes (see b) and d)) for density and absorbed power as a function of driving frequency.
    It has to be noted that panels b) and d) present data spanning over four orders of magnitude.
    Thus, a semi-logarithmic representation was chosen, and the following trends may not appear familiar to the reader's eyes.
    In Lieberman and Lichtenberg \cite{lieberman_principles_2005} one finds: $\bar{n}_\mathrm{e}\ \propto\ V_\mathrm{RF}$ for "low voltages", and $\bar{n}_\mathrm{e}\ \propto\ V_\mathrm{RF}^2$ for "high voltages".
    These relations explain the different trends in the region $V_\mathrm{RF} < 320\,$V and approximately define the first two discharge regimes regarding driving voltage.
    The first slope is visible for low voltages and lies inside the blue canvas (mode I, non-neutral).
    In a non-logarithmic representation, this slope has a linear trend.
    The second slope matches the high-voltage trend (i.e., in non-logarithmic representation, it shows quadratic behavior).
    This second characteristic is best visible for the $34\,$MHz cases, where the turning point between the first and second slopes is between $V_\mathrm{RF} \approx 180 - 200\,$V.
    There, all cases for $V_\mathrm{RF} = 220 \mathrm{\ to\ } 310\,$V show distinguished sheath and bulk structures (thus, they are in mode II, quasi-neutral).
    Furthermore, it has to be noted that the $12\,$MHz cases miss the second slope entirely.
    Raising the driving voltage at a low frequency is unsuitable for reaching quasi-neutrality.
    
    For higher voltages, both density and absorbed power jump to significantly higher values and then scale exponentially with the driving voltage ($\bar{n}_\mathrm{e} \propto \exp \left(V_\mathrm{RF} \right)$).
    This abrupt change in scaling marks the transition point to the constricted discharge.
    Above $V_\mathrm{RF} = 320\,$V, all cases jump abruptly to high densities and powers inside the green canvas (mode III, constricted).
    Constricted discharges exhibit a dc-like behavior and are close to the transition to thermal arc plasmas \cite{schroder_characteristics_2015, golda_comparison_2020, dunnbier_stability_2015, yang_comparison_2005, laimer_investigation_2005, laimer_glow_2006}.
    Thus, comparable to the dc discharge, a critical voltage (in the dc case the breakdown voltage) is necessary to initiate the constricted mode.
    Later, we will argue for the dc-like characteristics of constricted discharges.
    For now, the dc-like behavior explains why the vastly different densities and absorbed powers of the blue, brown, and green curves ($12$, $18$, and $34\,$MHz) almost merge as soon as the jump to the constricted mode occurs.
    Nevertheless, the observable drastic increase in the absorbed power is consistent with the drastic increase in gas temperature reported in other works \cite{winzer_rf-driven_2022, farouk_atmospheric_2008, dunnbier_stability_2015, golda_comparison_2020}.
    These works showed that a significant amount of the power input contributes to the device's heating.
    Thus, an increase in the jet's temperature correlates to a raised power input.
	
	\begin{figure}[t!]
		\centering
		\includegraphics[width=\textwidth]{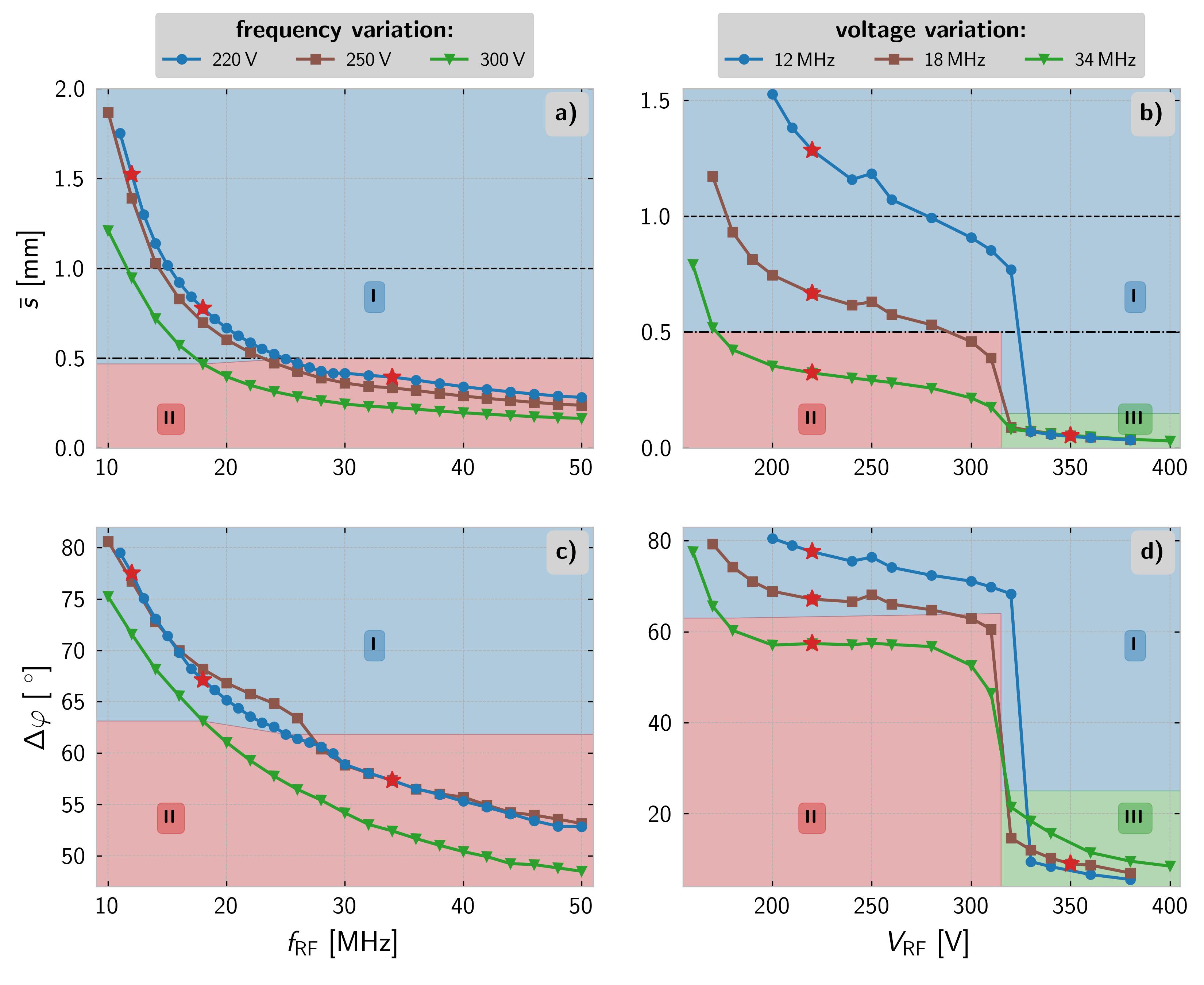}
		\caption{Theoretical mean sheath width $\bar{s}$ calculated based on an ion matrix sheath model \cite{lieberman_principles_2005, chabert_physics_nodate} as a function of driving frequency $f_\mathrm{RF}$ (a)) and voltage amplitude $V_\mathrm{RF}$ (b)) and phase shift $\Delta \varphi$ in the same dependency (c): frequency, d): voltage).
            The black horizontal lines in panels a) and b) mark the discharge gap (dashed) and half the discharge gap (dash-dotted).
            The colored polygonic shapes in the background approximately mark where each panel's operation modes are located.
            I: non-neutral mode (blue), II: quasi-neutral mode (red), and III: constricted mode (green).
            The red stars mark the cases where sections \ref{Dynamics1} and \ref{Dynamics2} will provide deeper analysis.
            All cases share the same base parameters: $p = 101325\,$Pa, $x_\mathrm{N_2} = 0.001$, $T_\mathrm{g} = 300\,$K.}
		\label{fig:sheath_width}
	\end{figure}

    Based on the different scaling for voltages $V_\mathrm{RF}<320\,$V, the presented assumption that "low voltages" coincide with non-neutral discharges while "high voltages" mark the region where quasi-neutral plasmas were found stands.
    However, the terms "low" and "high" voltages are ambiguous (especially when "high" means intermediate voltages below the critical high voltage for constriction).
    Moreover, doing a full voltage sweep (that potentially breaks the jet) proves slightly impractical in determining the operation mode.
	Thus, the question remains: How do we differentiate between the discharge modes?

    Figure \ref{fig:sheath_width} provides two means of answering the question.
    The figure shows the mean sheath width $\bar{s}$ in the top row and the phase shift between current and voltage $\Delta \varphi$ in the bottom row.
    The left column shows the data for a frequency variation.
    For the right column, $\bar{s}$ and $\Delta \varphi$ are shown as a function of the driving voltage $V_\mathrm{RF}$.
    Similar to figure \ref{fig:parameters}, the color of the canvas gives an orientation about the discharge mode (blue = I non-neutral mode, red = II quasi-neutral mode, green = III constricted mode).
    The frequency variation (left) has been conducted for three different voltages (blue = $220\,$V, brown = $250\,$V, green = $300\,$V).
    When varying the voltage (right), three different frequencies have been used (blue = $12\,$MHz, brown = $18\,$MHz, green = $34\,$MHz).

    Calculating the sheath width is generally a tedious task \cite{brinkmann_electric_2015, naggary_bridging_2019, klich_validation_2022}.
    It becomes especially challenging for non-neutral discharges when there is no plasma bulk \cite{klich_simulation_2022}.
    Thus, we opt for calculating a theoretical mean sheath width $\bar{s}$ by taking the mean total ion density $\bar{n}_\mathrm{i,tot} = \sum_s \bar{n}_{\mathrm{i,}s}$ of every case as an input value for a simplistic ion matrix sheath model \cite{lieberman_principles_2005, chabert_physics_nodate}.
    The matrix sheath model calculates the mean sheath width $\bar{s}$ by the formula $\bar{s} = \sqrt{\frac{2\, \varepsilon_0\, V_\mathrm{RF}}{e\, \bar{n}_\mathrm{i,tot}}}$.
    In this formula, $\varepsilon_0$ is the electric field constant, and $e$ is the elementary charge.
    This representation of a sheath width might not be the most accurate.
    However, evaluating this formula for the most extreme cases (i.e., the non-neutral ones) produces a reference value for comparing all discharges.
    For further reference, the dashed line in panels a) and b) refers to the discharge gap of $1\,$mm, and the dash-dotted line marks half of it, the upper limit for forming a steady plasma bulk.
    By this reference, it becomes evident that for the highly non-neutral cases (in panel a) $f_\mathrm{RF} \approx 10\,$MHz, and in panel b) most $12\,$MHz cases) not even a single sheath fits inside the discharge gap.
    Thus, the discharge cannot shield a bulk region from the applied voltage, and the electrons react by possessing unique dynamics discussed previously in \cite{klich_simulation_2022}.
    Many of the simulated cases end up in an area with a blue background.

    Both the frequency and the voltage variation show similar trends.
    Increasing the frequency and voltage causes the plasma density to rise and the boundary sheath width to shrink \cite{lieberman_principles_2005, chabert_physics_nodate}.
    Thus, there is a smooth transition from non-neutral discharges to quasi-neutral plasmas.
    When, on average, one sheath fits in the discharge, a temporally and spatially highly modulated plasma bulk forms.
    This bulk region stabilizes when the plasma becomes dense enough for two sheaths to fit inside the discharge gap.
    However, it has to be stressed that there is a significant difference between raising the driving frequency and increasing the driving voltage.
    The plasmas simulated in this work cannot enter the constricted mode by choosing a fixed driving voltage (underneath the critical voltage $V_\mathrm{c} \approx 320\,$V) and varying the frequency.
    Thus, solely the voltage variation can trigger the abrupt transition to the constricted mode (III).
    
    Moreover, panel a) shows that discharges for frequencies below $18\,$MHz do not show quasi-neutral behavior for the considered range of voltages (i.e., $220$, $250$, and $300\,$V).
    Panel b) additionally supports that our simulation never reaches the quasi-neutral mode for $12\,$MHz.
    Two boundary sheaths never fit the discharge gap for $V_\mathrm{RF} < 330\,$V.
    For higher voltages ($V_\mathrm{RF} > 330\,$V), the threshold voltage is surpassed and the discharge becomes constricted.
    Regarding the average sheath width, the constriction becomes visible by $\bar{s}$ sharply dropping to values below 1/8\textsuperscript{th} of the discharge gap.

    While determining the particle density in a micro-discharge is rather tricky \cite{bruggeman_foundations_2017}, the phase shift $\Delta \varphi$ is experimentally more accessible.
    Panels c) and d), the phase shift is discussed as a possible experimental method for determining the discharge mode.
    Both show that for the limiting case of a highly non-neutral case (i.e., a capacitor filled with a non-distorted positive charge density), the phase shift tends towards $90^\circ$.
    When the plasma density increases, the sheaths shrink, and the discharge becomes increasingly quasi-neutral; the plasma becomes more resistive.
    This trend of quasi-neutral plasmas at elevated pressures being increasingly resistive is in accordance with previous work \cite{vass_evolution_2022}.
    Moreover, the phase-shift plateaus at some value characteristic for the given frequency and voltage (e.g., $f_\mathrm{RF} > 30\,$MHz \@ $220\,$V in panel c) or $V_\mathrm{RF} \approx 200 - 300\,$V \@ $34\,$MHz in panel d)).
    This plateau appears for quasi-neutral discharges and marks the presence of a steady plasma bulk region.
    The transition to the constricted mode also becomes visible in panel d) by a jump in the phase shift.
    Yet again, this jump solely occurs when a critical voltage ($V_\mathrm{c} \approx 320\,$V) is surpassed.
    Thus, the frequency variation in panel c) lacks this feature.
    The dc-like characteristics of constricted plasmas include an almost purely resistive behavior ($\Delta \varphi \approx 5 - 22^\circ$).
    
    Once more, the phase shift reveals that the $12\,$MHz cannot form a quasi-neutral plasma (s. panel d)).
    The phase shift does not show the characteristic plateau and stays at high values for capacitive loads ($\Delta \varphi > 70^\circ$) before transitioning rapidly towards the constricted mode.
    The low phase shift supports the finding of previous studies that made a connection between constricted and dc discharges \cite{dunnbier_stability_2015, farouk_atmospheric_2008}.

    This section has shown three distinct discharge modes (i.e., non-neutral, quasi-neutral, and constricted).
    The non-neutral and quasi-neutral modes can be smoothly switched between by raising the plasma density by both applied means (voltage and frequency).
    The transition to the constricted mode is abrupt and requires a critical voltage ($V_\mathrm{c} \approx 320\,$V) to be applied.
    Thus, raising the driving frequency of the simulation was not a sufficient tool to generate a constricted plasma.

%%%%%
%%%%%
%%%%%
\subsection{Experimental data}\label{exp_data}
    \begin{figure}[t!]
        \centering
        \includegraphics[width=\textwidth]{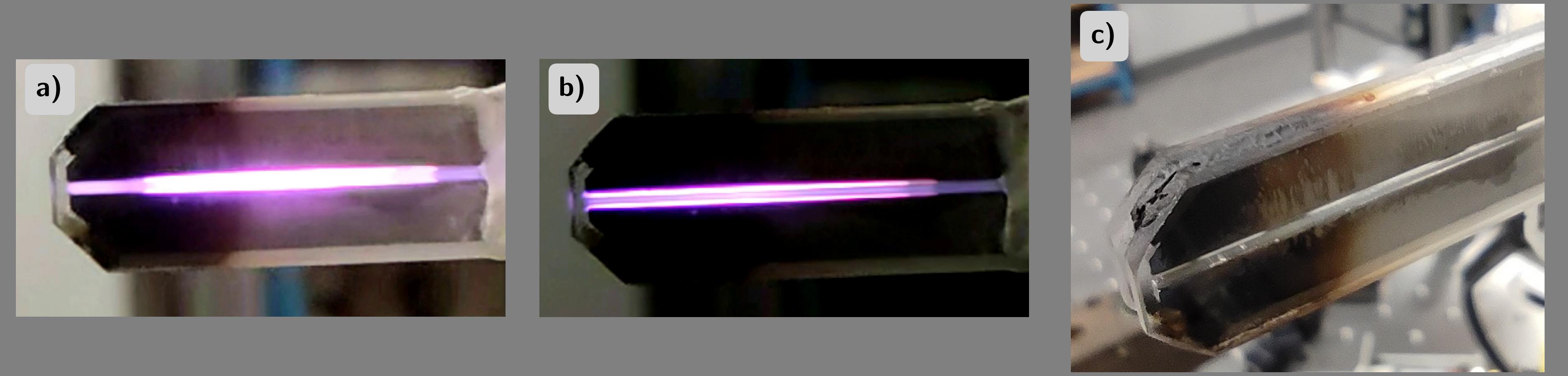}
        \caption{Pictures of the COST-Jet operated in the constricted mode.
        (a)) A photograph of the constricted mode as seen by the naked eye, (b)) the same experiment conducted with an adjusted camera shutter speed to avoid image saturation, and (c)) Damaged COST-Jet after a few tens of seconds of operation.}
        \label{fig:experiment}
    \end{figure}
    
    In previous work, Golda et al. showed that a constricted plasma could be ignited inside the COST-Jet \cite{golda_comparison_2020}.
    However, this constricted plasma was localized close to the nozzle and tended to transit to a thermal arc plasma \cite{laimer_investigation_2005} that eventually damaged the jet.
    The excessive heat production of constricted plasmas and their tendency to thermalize was reported by other authors as well \cite{farouk_atmospheric_2008, dunnbier_stability_2015}.
    The constricted mode presented in this work is similar to that observed in studies by Yang et al. \cite{yang_comparison_2005} and Laimer et al. \cite{laimer_glow_2006, laimer_investigation_2005}.
    Both groups show constricted plasmas in stable operation.
    Like the pictures shown in figure \ref{fig:experiment}, the plasma is not localized to a spot but spread out.
    In our case, by delivering more power, the constricted discharge fills almost the entire length of the COST-Jet's discharge channel.
    In panel a), the plasma's constriction is visible through the intense light emission; the camera shutter speed was adjusted to capture this image as seen by the naked eye.
    By increasing the camera shutter speed to avoid image saturation (panel b), a bright light is observed in the vicinity of the electrode, and the less bright positive-column-like bulk region is seen in between.

    For the present experiment, we can confirm the intense heat production of the constricted plasma mentioned above \cite{farouk_atmospheric_2008, laimer_glow_2006, golda_comparison_2020}.
    As shown in figure \ref{fig:parameters} (d), the constricted mode is a high-power operation mode.
    Kelly et al. \cite{kelly_gas_2015} demonstrated that the power input is partially converted into carrier gas heating.
    We observe that during the operation of the jet in the constricted mode, the electrodes become very hot. 
    The Torr Seal$^{\circledR}$ glue used to assemble the device \cite{golda_concepts_2016} is, however, not well-suited for high-power operation, as it has an operational temperature range of -45$^\circ$C - 120$^\circ$C. 
    Panel c) shows a close-up photograph of a COST-Jet that got severely damaged by excessive heat in less than a few tens of seconds during the attempts to get stable discharge conditions for the PROES measurements.

    \begin{figure}[t!]
        \centering
        \includegraphics[width=\textwidth]{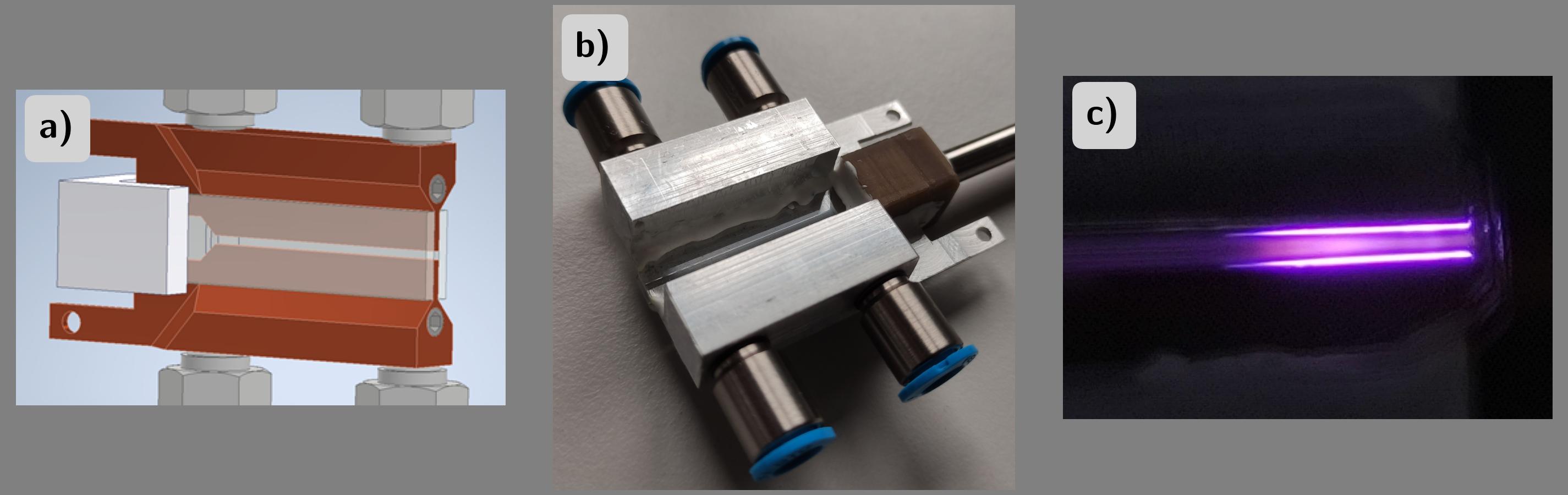}
        \caption{The improved version of the COST-Jet with cooled electrodes.
            Panel a) shows the 3D CAD design of the modified jet.
            Panel b) presents the prototype of the modified COST-Jet.
            Panel c): Photograph showing the constricted mode ignited in the cooled jet using helium with a 0.1\% nitrogen admixture: $f_\mathrm{RF} = 13.56\,$MHz and $V_\mathrm{RF} = 350\,$V.}
        \label{fig:exp_cooled}
    \end{figure}

    We attempted to operate the jet in a pulsed mode to overcome the issue of excessive heat production and prevent jet destruction during measurements.
    During the discharge-off phase, we set all necessary parameters for the measurements.
    During the discharge-on phase, we performed camera measurements at specific times relative to the RF period.
    We repeated the process until the entire RF period was covered to obtain excitation plots.
    Unfortunately, despite varying the pulse frequency and duty cycle, we could not achieve reproducible or valuable results.
    Therefore, the COST-Jet's design had to be modified to introduce additional cooling to the electrodes and prevent damage to glued areas.

    To keep the COST-Jet design as close as possible to the original construction, the original electrodes were extended away from the discharge area, and water cooling was introduced for the extended part.
    Figure \ref{fig:exp_cooled} a) shows a CAD construction sketch of a modified version of the COST-Jet. 
    The design allows the cooling of the electrodes using either compressed air or water.
    The holes on the electrodes' expanded part are kept open for compressed air cooling.
    These holes are closed for water cooling, with water entering from the left pipe and exiting through the right one.
    Since one of the electrodes is powered in this experiment, compressed air was used for both electrodes for safety measures.
    The prototype of this cooled jet is shown in panel b), and the resulting constricted plasma is visible in panel c). 
    
    \begin{figure}[t!]
        \centering
        \includegraphics[width=\textwidth]{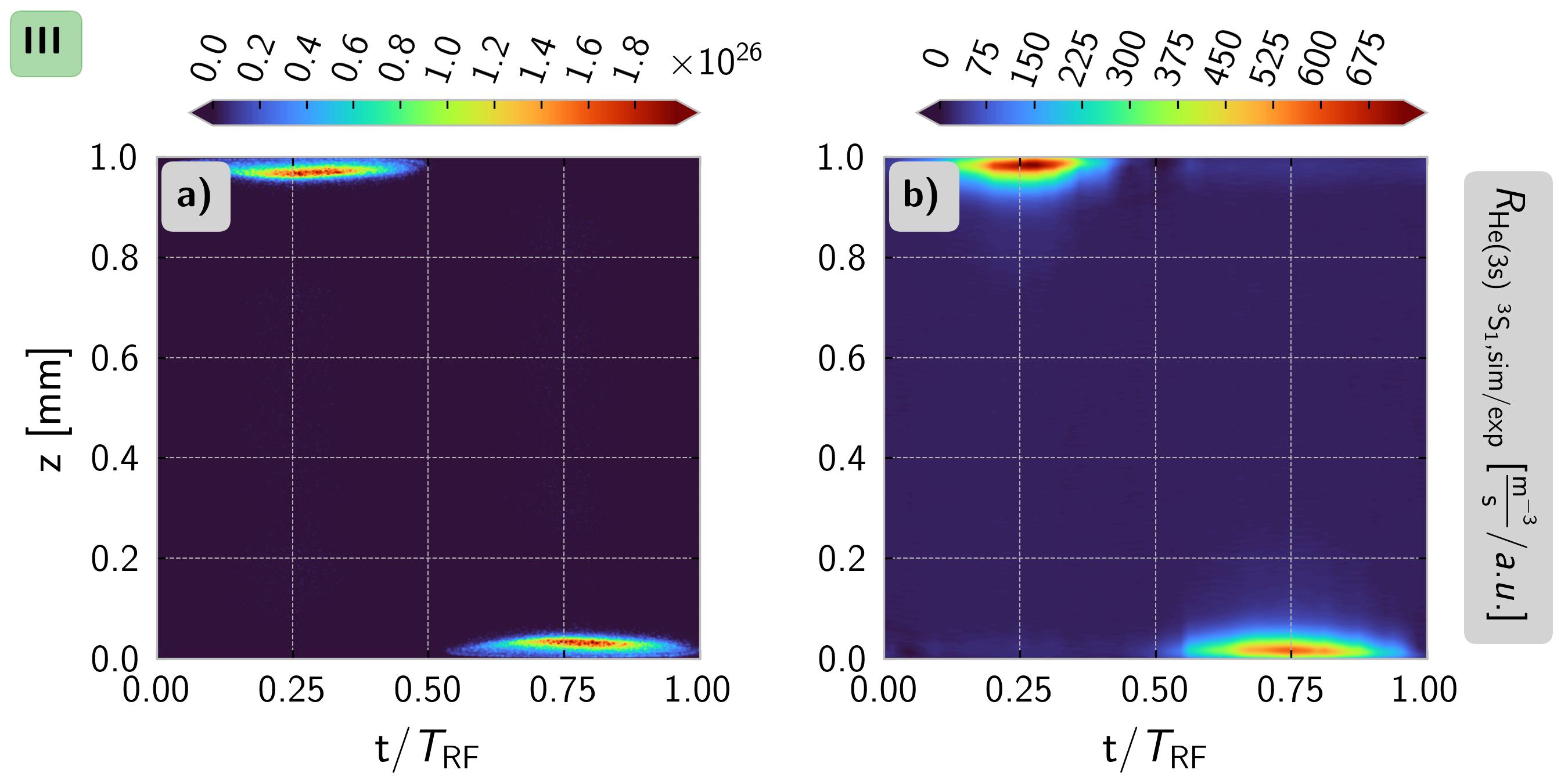}
        \caption{Spatio-temporal distributions of the electron impact excitation rate from the ground state into He-I (3s) $^3$S$_1$  Panel a) simulation data, panel b) experimental data obtained by PROES measurement by monitoring the He line at $706.5\,$nm. Color scale in arbitrary units (a.u.).
            Common parameters of simulation and experiment: $f_\mathrm{RF} = 13.56\,$MHz, $x_\mathrm{N_2} = 0.001$.
            Conditions of the simulations: $V_\mathrm{RF} = 340\,$V, $p = 101325\,$Pa, $T_\mathrm{g} = 300\,$K.
            Conditions of the experiment: $V_\mathrm{RF} = 350\,$V.}
        \label{fig:PROES}
    \end{figure}

    Despite these modifications, after approximately 5 minutes of operation in the constricted mode, the power of the jet fluctuates between 10W and 15W, and we also observe notable changes in the transparency of the quartz glass.
    Nevertheless, the new construction allows us to perform relatively stable PROES measurements before these changes occur.
    
    The results of the measurements are shown in figure \ref{fig:PROES}.
    The figure compares the spatio-temporal distributions of the computed (panel a)) and obtained experimentally (panel b) electron impact excitation rate from the ground state into He-I (3s) $^3$S$_1$.
    There is an excellent qualitative agreement between the measured and computed excitation rates. 
    Both the experiment and the simulation show the maximum excitation rate confined to the respective momentary anode (first half-wave upper electrode, second half-wave lower electrode).
    Simultaneously, the excitation pattern sinusoidally grows and shrinks with the sheath expansion.
    The slight asymmetry in the maximum excitation rate between the first and second halves of the RF period is attributed to the slightly changing experimental conditions during the measurements mentioned above (c.f. panel b)).
    
    Since measurements in this mode are pretty challenging and the main aim of this paper is to investigate this mode using simulations, we have limited ourselves to this one set of measurements to confirm our simulation findings.
    Further experimental investigations will require more careful jet design, exploration of different conditions and power levels, and analysis of the effects on the transparency of the quartz plates after extended operation in this mode.
    
%%%%%
%%%%%
%%%%%
 \subsection{Discharge dynamics of non-neutral and quasi-neutral discharges}\label{Dynamics1}
    \begin{figure}[t!]
        \centering
        \includegraphics[width=\textwidth]{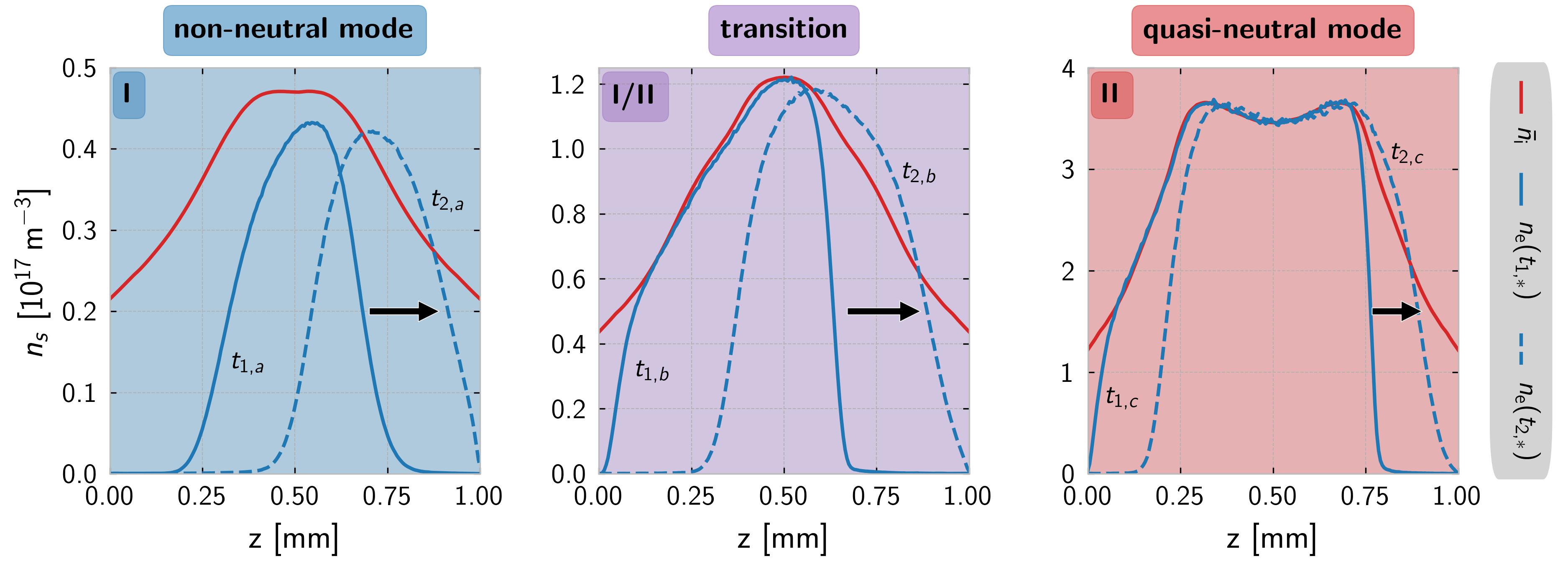}
        \caption{Snapshots of spatially resolved density profiles.
            The ion density $\bar{n_\mathrm{i}}$ is shown on time-average.
            Electrons $n_\mathrm{e}$ are shown when they are located approximately in the center of the discharge gap ($t = t_{1,\ast}$) and during the sheath collapse ($t = t_{2,\ast}$).
            Panel I: non-neutral discharge ($f_\mathrm{RF} = 12\,$MHz), I/II: transition between regime I and II ($f_\mathrm{RF} = 18\,$MHz, and III: quasi-neutral mode ($f_\mathrm{RF} = 34\,$MHz).
            All cases share the same base parameters: $V_\mathrm{RF} = 220\,$V, $p = 101325\,$Pa, $x_\mathrm{N_2} = 0.01$, $T_\mathrm{g} = 300\,$K.}
        \label{fig:snapshots}
    \end{figure}

    Section \ref{general} described the time-averaged density profiles of three distinct discharge modes and the transition between them using global (i.e., time- and space-averaged) data.
    This section will dive into the spatio-temporal dynamics of non-neutral and quasi-neutral discharges.
    Since it was shown that the non-neutral mode and the quasi-neutral mode can appear for both a variation of the driving voltage and frequency, we now focus on varying the frequency.
    In the following, the dynamics of three cases are analyzed operated at $V_\mathrm{RF} = 220\,$V and different frequencies ($12\,$MHz for the non-neutral case, $18\,$MHz for a case at the smooth transition, and $34\,$MHz for the quasi-neutral case).
    The transition between regime I and II is accordingly labeled I/II.
    
    The starting point of this discussion is displayed in figure \ref{fig:snapshots}.
    The figure shows spatially resolved density profiles in the panels (allocation described above).
    These density profiles are snapshots of the electron (blue) and total ion density (red).
    The ions are almost non-modulated in time and therefore shown as static by their mean value $\bar{n}_\mathrm{i}$.
    The electrons are shown at two times within the RF period.
    The first time $t_\mathrm{1,\ast}$ is at the beginning of the sheath expansion.
    (Please note that the use of the term "sheath" is meant in a figurative sense for non-neutral discharges and the transition regime, cf. section \ref{general}.)
    At this time, the electrons of the non-neutral case (panel I) are organized around the center of the discharge.
    Additionally, this time generates the largest expansion of the quasi-neutral region of the transition case (panel I/II).
    The second time $t_\mathrm{2,\ast}$ is chosen to best display a particular feature of these high-pressure discharges.
    During the sheath collapse, electrons always overshoot the total ion density and create a local negative space charge in the process.
    The comparison of panels I and II shows that this characteristic happens regardless of the discharge mode.
    The following section will prove that this statement also applies to the constricted mode.
    (Animations of figure \ref{fig:snapshots}'s data are provided with the accompanying material online.)
    
	\begin{figure}[t!]
		\centering
		\includegraphics[width=\textwidth]{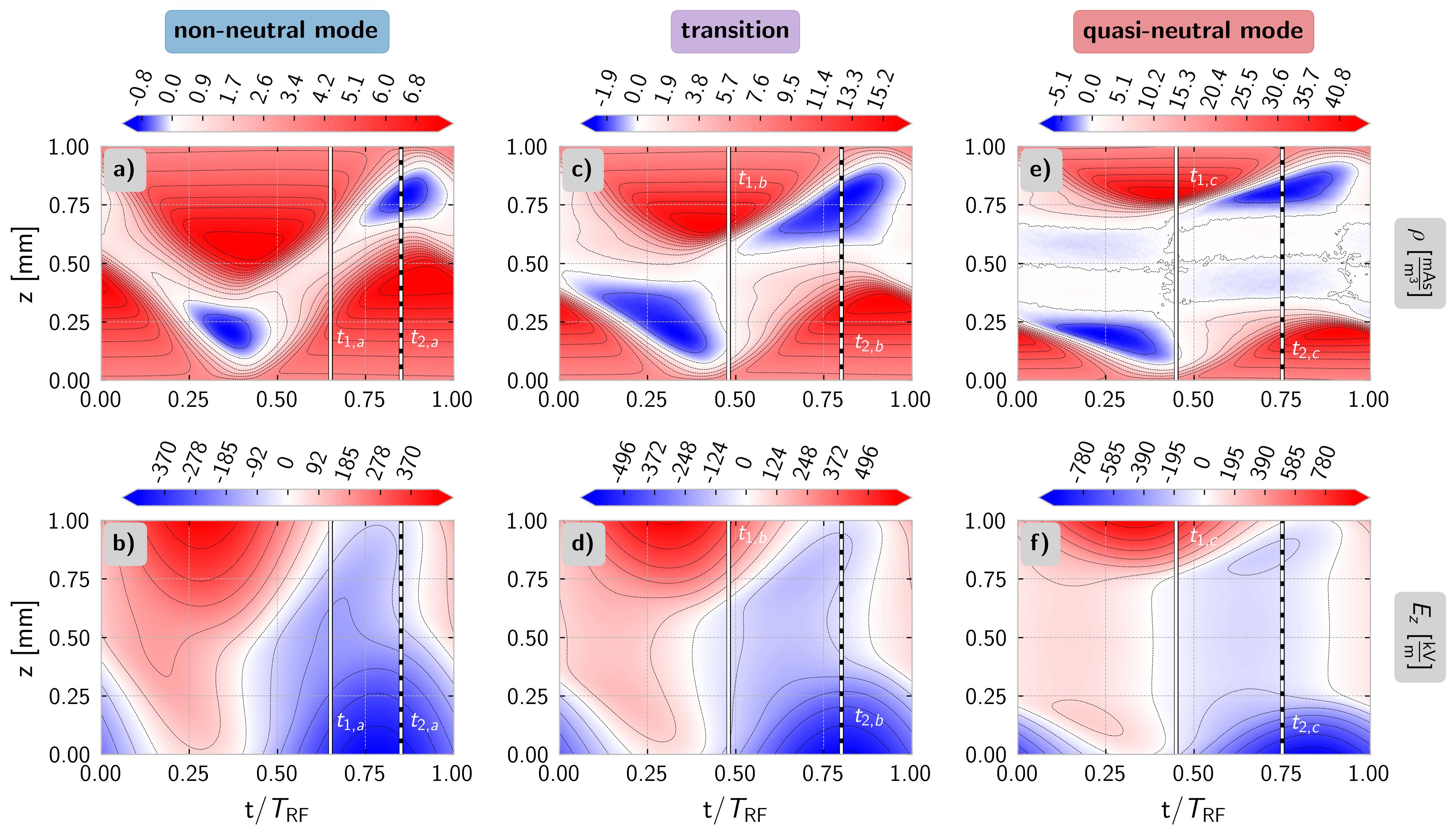}
		\caption{Time and space resolved charge density $\rho$ (top row) and electric field $E_z$ (bottom) for a non-neutral discharge (left column, $f_\mathrm{RF} = 12$MHz), a discharge in transition (second column, $f_\mathrm{RF} = 18$MHz), and a quasi-neutral discharge (right column, $f_\mathrm{RF} = 34$MHz).
            The solid and dashed white lines mark the temporal positions $t_{1,\ast}$ and $t_{2,\ast}$ of the snapshots of figure \ref{fig:snapshots} for reference.
            All cases share the same base parameters: $V_\mathrm{RF} = 220\,$V, $p = 101325\,$Pa, $x_\mathrm{N_2} = 0.01$, $T_\mathrm{g} = 300\,$K.}
		\label{fig:charge_density}
	\end{figure}

    To further investigate the plasma dynamics connected to this overshooting of electrons, figure \ref{fig:charge_density} shows spatially and temporally resolved data for the non-neutral mode (left), the transition regime (middle), and the quasi-neutral mode (right).
    The top row shows the cases' charge density $\rho$, and the bottom row has the related electric fields $E_z$.
    For reference, the times $t_\mathrm{1,\ast}$ (bold) and $t_\mathrm{2,\ast}$ (dashed) of the snapshots in figure \ref{fig:snapshots} are marked by white lines.
    In general, figure \ref{fig:charge_density} shows the development of charge density and electric field during the formation of a steady plasma bulk.
    Panels a) and b) show the non-neutral situation without any plasma bulk.
    In panels c) and d), a tiny quasi-neutral bulk region oscillates around $z=0.5\,$mm.
    And, for the quasi-neutral scenario ((e) and f)), there is a steady plasma bulk from $z \approx 0.3$ to $0.7\,$mm that distinctly separates sheath and bulk dynamics.

    Although the features described above lead to individual dynamics, all three cases share a common characteristic.
    As discussed in figure \ref{fig:snapshots}, electrons overshoot the ion density when driven towards the electrode during the sheath collapse.
    This overshooting of the electrons is visible in the charge density as a negative charge density at $t \approx t_\mathrm{2,\ast}$.
    Depending on the discharge conditions, these regions of negative charge density vary in size and intensity.
    For non-neutral discharges (panel a)), the negative space charge region is the smallest and weakest.
    Whereas for quasi-neutral plasmas (panel e)), the space charge is strongest and has the broadest temporal spread.
    Generally, the more quasi-neutral a discharge becomes, the higher the density and the higher the uncompensated charge caused by overshooting electrons.
    Furthermore, the spatial stretch of the negative space charge grows during the transition from non-neutral to quasi-neutral yet shrinks again when the bulk region enlarges (c.f. top row).
    Additionally, the temporal position of the sheath collapses, and thus, the moment electrons begin to overshoot the ion density shifts to earlier moments in the RF cycle.

    This observation is connected to the decrease in the phase shift (cf. section \ref{general}).
    The more quasi-neutral the discharge, the lower the phase shift $\Delta \varphi$.
    Consequently, the current extremes occur earlier (compared to discharges closer to the non-neutral mode).
    These times of highest current are linked to the complete collapse of the boundary sheath.
    Thus, sheath expansion also happens earlier (and $t_\mathrm{1,\ast}$ moves to earlier times too).
    Moreover, the need to drive the electron current through the discharge (to maintain charge neutrality) causes the electrons to overshoot the ion density.
    Electrons at high pressures suffer from many collisions; thus, much friction opposes the electron's movement.
    In situations like this, and especially at high pressures, it is known that a field reversal is most likely to form \cite{czarnetzki_space_1999, mohr_field_2013}.
    The lower row of figure \ref{fig:charge_density} shows that regardless of the discharge mode, the overshoot of electrons and the negative space charge coincide with a field reversal directly in front of the electrode.
    At and around $t = t_\mathrm{2,\ast}$, an electric field in front of the upper electrode accelerates electrons towards the electrode.
    This field reversal is less noticeable in the characteristic uni-polar field structure of non-neutral discharges \cite{klich_simulation_2022} (c.f., panel b)).
    Yet, it evolves into distinct local extremes for quasi-neutral discharges (c.f. panel d) and f)).
    Nevertheless, electrons are accelerated towards the respective electrodes within these field reversals.
    The same dynamics happen at the mirrored times ($t \approx 0.2 - 0.25$) at the lower electrode.
    
    \begin{figure}[t!]
		\centering
		\includegraphics[width=\textwidth]{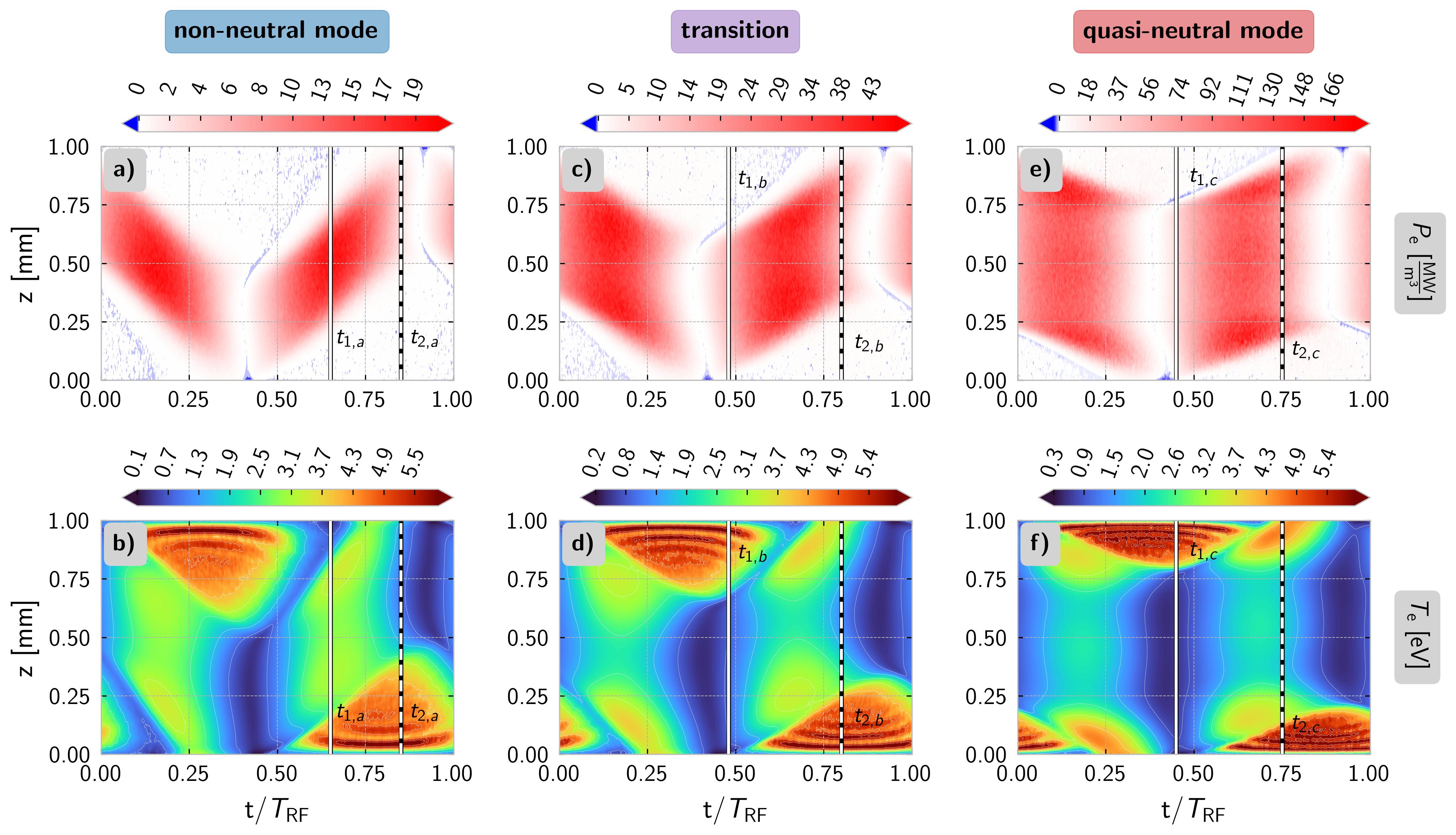}
		\caption{Spatially and temporally resolved electron power density $P_\mathrm{e}$ (top row) and electron temperature $T_{\mathrm{e}}$ (bottom row) for a non-neutral (12 MHz, left column), a quasi-neutral (34MHz, right column) and a discharge in between (18MHz, middle column).
            The solid and dashed white lines mark the temporal positions $t_{1,\ast}$ and $t_{2,\ast}$ of the snapshots of figure \ref{fig:snapshots} for reference.
            All cases share the same base parameters: $V_\mathrm{RF} = 220\,$V, $p = 101325\,$Pa, $x_\mathrm{N_2} = 0.01$, $T_\mathrm{g} = 300\,$K.} 
		\label{fig:power_density}
	\end{figure}

    We have now seen that electrons overshoot the ion density and create negative space charges.
    Moreover, electric field reversals are observed in two of three operation modes.
    The following will show that the discharge modes possess common dynamics based on the field reversal structure.
    Figure \ref{fig:power_density} presents the spatio-temporal dynamics of the power absorption for the three cases (left: non-neutral, mid: transition, right: quasi-neutral).
    The power absorption dynamics are represented by the electron power density $P_\mathrm{e} = \vec{j}_e \cdot \vec{E}$ (upper row) and the electron temperature $T_\mathrm{e}$ (lower row).
    As in previous work, the electron temperature is calculated on a kinetic basis and extracted from the diagonal element $p_{zz}$ of the pressure tensor \cite{wilczek_electron_2020, klich_simulation_2022, klich_validation_2022}.

    Previous work  \cite{klich_simulation_2022} has extensively analyzed the power dynamics of the non-neutral mode (left).
    We describe there that in non-neutral discharges, the plasma dynamics center around the electrons grouping up and following a midpoint in Boltzmann equilibrium.
    Moreover, previous work has shown that the power absorption is almost exclusively ohmic \cite{vass_electron_2021}.
    Nevertheless, revisiting the power absorption dynamics is worthwhile.

    The basic logic of the power dynamics of non-neutral plasmas is that, due to the high friction, power is dissipated whenever the electron group moves.
    A similar pattern is observable for all the discharges shown in figure \ref{fig:power_density}.
    Power is dissipated where electrons are most volatile.
    Following this reasoning, there are two local hot spots where many electrons are present and under the influence of strong electric fields.
    \begin{enumerate}
        \item In front of the expanding sheath.
            Panels c) and e) show each a local maximum of the electron power density $P_\mathrm{e}$ in front of the expanding sheath.
            This maximum occurs at the lower electrode between $t_\mathrm{1,b/c}$ and $t_\mathrm{2,b/c}$.
            The corresponding electron temperature $T_\mathrm{e}$ in panels d) and f) also contains local maxima in the same spatio-temporal positions.
            This peak is the $\Omega$-peak characteristic for the $\Omega$-mode \cite{hemke_ionization_2012, bischoff_experimental_2018, korolov_control_2019, korolov_helium_2020, schulenberg_mode_2024} and often described as $\alpha$-peak due to its similarity to the $\alpha$-mode in low pressure \cite{golda_comparison_2020, dunnbier_stability_2015, schulz-von_der_gathen_spatially_2008}.
            Additionally, a close consideration of panel b) reveals that a similar but less intense structure is also found in the non-neutral mode.
        \item The second hot spot is during the field reversal, meaning in front of the collapsing sheath.
            Due to the high pressure and, accordingly, friction, electrons need to be accelerated out of the discharge to leave in time before the sheath expands again.
            This effect leads to a field reversal, a local extreme of the electric field in quasi-neutral plasmas, which causes electrons to gain energy.
            All plots show that this energy gain happens in the same time frame $t_\mathrm{2,\ast}$ (i.e., when electrons overshoot the ion density).
            This energy gain is visible in the electron power density and temperature (c.f. panels c) -- f)).
            Panel b) shows that this is again less bright than in the other cases but is also visible for the non-neutral case.
    \end{enumerate}

    Overall, the absence of two distinguishable peaks in the electron power density of the non-neutral case in panel a) is due to the missing bulk structure separating the peaks.
    Without a plasma bulk, the $\Omega$- and the field-reversal-peak merge for the power density.
    Both structures grow apart with a growing plasma bulk and eventually form distinct local maxima.
	
	\begin{figure}[t!]
		\centering
 		\includegraphics[width=\textwidth]{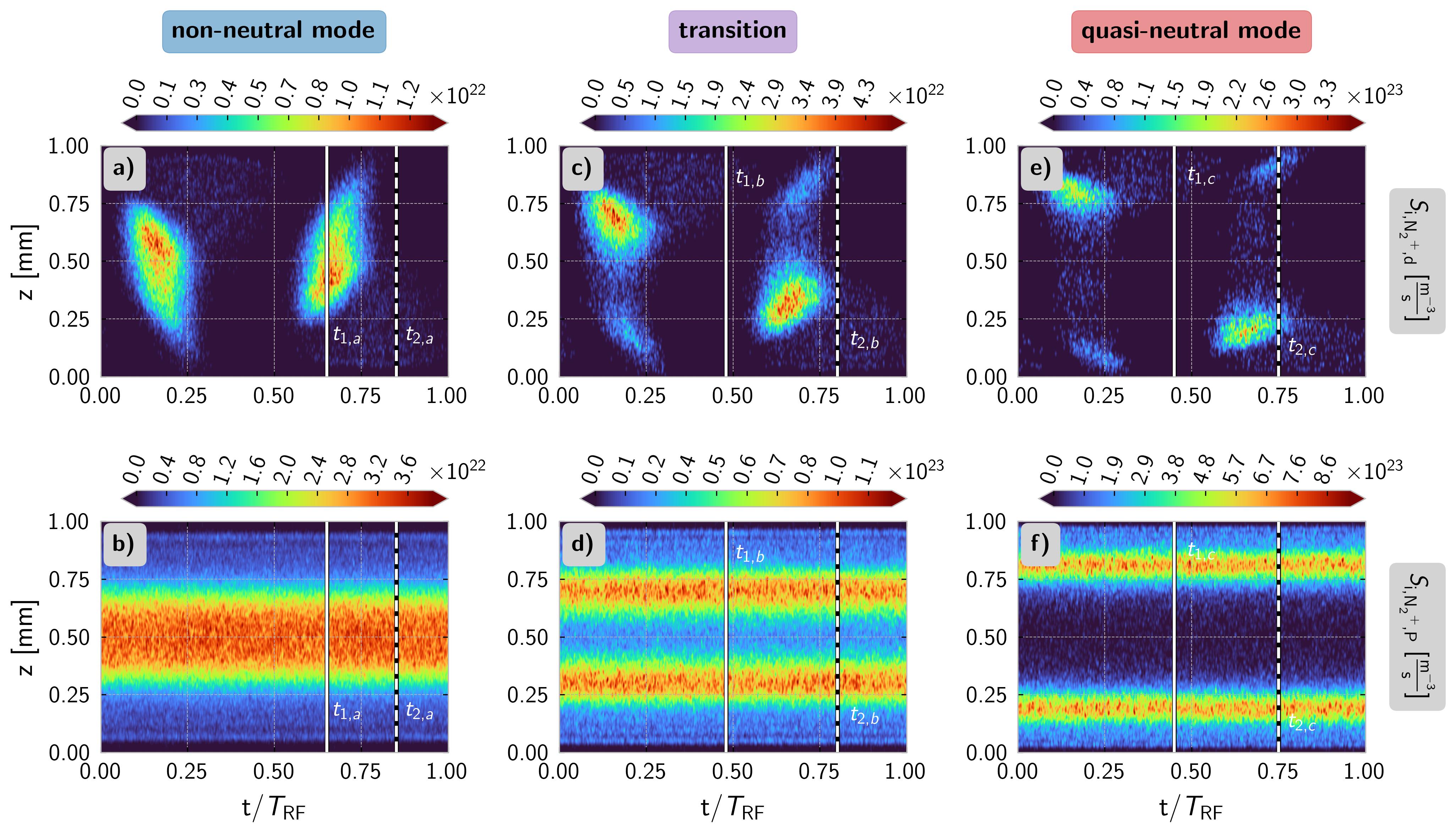}
		\caption{Spatially and temporally resolved nitrogen ionization dynamics for a non-neutral discharge ($f_\mathrm{RF} = 12\,$MHz, left column), a discharge in transition ($f_\mathrm{RF} = 18\,$MHz, middle column), and a quasi-neutral discharge ($f_\mathrm{RF} = 34\,$MHz, right column).
            The ionization is separated into electron impact ionization $S_\mathrm{i, N_2^{\ +},d}$ (top row) and the Penning ionization $S_\mathrm{i, N_2^{\ +},P}$ (bottom row).
            The solid and dashed white lines mark the temporal positions $t_{1,\ast}$ and $t_{2,\ast}$ of the snapshots of figure \ref{fig:snapshots} for reference.
            All cases share the same base parameters: $V_\mathrm{RF} = 220\,$V, $p = 101325\,$Pa, $x_\mathrm{N_2} = 0.01$, $T_\mathrm{g} = 300\,$K.} 
		\label{fig:ion_dynamics}
	\end{figure}

    The ionization dynamics shown in figure \ref{fig:ion_dynamics} reflect the same trend and features analyzed for the electron power dynamics.
    Figure \ref{fig:ion_dynamics} again shows the first two operation modes and their transition, each by one column (left: non-neutral, mid: transition, right: quasi-neutral).
    The shown spatially and temporally resolved nitrogen ionization profiles are separated into two parts:
    (i) Electron impact ionization of N\textsubscript{2} in the top row,
    (ii) and Penning ionization of N\textsubscript{2} (s. 31 in tab. \ref{tab:chemistry}) in the bottom row.

    The electron impact ionization (top row) mimics the corresponding electron temperatures.
    For non-neutral discharges, ionization occurs when the electron group is moving (compare fig. \ref{fig:power_density} b) to \ref{fig:ion_dynamics} a)) \cite{klich_simulation_2022}.
    With increasing density and, accordingly, quasi-neutrality of the discharge, the ionization maxima (i.e., the $\Omega$- and the field-reversal-peak) become distinguishable and separated by the growing plasma bulk (cf. fig. \ref{fig:ion_dynamics} top row).
    The observation of two distinct ionization structures and the interpretation of their origin is analog to the emission structures observed by Scharper et al. \cite{schaper_electron_2011} and Gibson et al. \cite{gibson_disrupting_2019}.
    
    The temporal structure of the Penning ionization is diffuse.
    Nevertheless, the data show a spatial trend comparable to previously discussed.
    Non-neutral cases (panel b)) have one broad maximum in the center of the discharge, and the spatial profile is similar to a diffusion profile.
    With increasing quasi-neutrality (c.f. d) and f)), the single peak splits into a bimodal structure with maxima that move close towards the electrodes since the plasma bulk becomes less energetic and power dissipation is tied to the interaction with shrinking boundary sheaths (c.f. fig. \ref{fig:power_density}).

    Overall, the comparison and analysis of the electron dynamics during the transition of non-neutral discharges (mode I) to quasi-neutral plasmas (mode II) has shown that there is a common element.
    Due to the high pressure, electrons always overshoot the ion density when driven towards an electrode.
    This overshoot coincides with a field reversal in which electrons gain considerable energy.
    As a consequence of the energy gain, the power and ionization dynamics show a distinct field-reversal peak.
    The field interactions (i.e., expanding sheath and field reversal) are barely distinguishable in the non-neutral mode.
    With growing plasma density and, accordingly, quasi-neutrality, these structures become separable.
    
%%%%%
%%%%%
%%%%%
\subsection{Dynamics of the constricted mode}\label{Dynamics2}

    Variation of either the frequency or the driving voltage can achieve the transition and dynamics discussed in the previous section.
    However, the constricted mode requires a critical field strength (and thus a minimal voltage).
    Therefore, only voltage variation can produce plasmas in the constricted mode.
    This section will analyze the dynamics of the constricted mode (III) and present analogies to the previous two modes.
    For doing so, an $18\,$MHz case at $V_\mathrm{RF} = 350\,$V is arbitrarily chosen from the pool of simulated cases in the constricted mode.
    
    As discussed in previous work \cite{farouk_atmospheric_2008, zhang_excitation_2009, dunnbier_stability_2015, yang_comparison_2005, laimer_glow_2006, laimer_investigation_2005}, the dynamics of constricted cases focus on narrow regions directly in front of the electrode.
    These regions are the tiny sheaths that cause enormous electric fields to occur and centralize the main dynamics to them.
    Hence, constricted plasmas are conceived as a kind of high-pressure $\gamma$-mode \cite{dunnbier_stability_2015, zhang_excitation_2009, laimer_glow_2006, laimer_investigation_2005}.
    However, overshadowed by the phenomena in the boundary sheaths, the same dynamics as discussed before are still present and will be identified in the following.

    Analog to the analysis of the dynamics of modes I and II, we will start at the densities and field and end up with the ionization. 
    In contrast to the previous representations, the following plots show only half of the discharge gap.
    Due to the discharge's symmetry, no crucial information is lost.
    Yet, the minuscule structures of interest are better visible.
		
    \begin{figure}[t!]
        \centering
        \includegraphics[width=\textwidth]{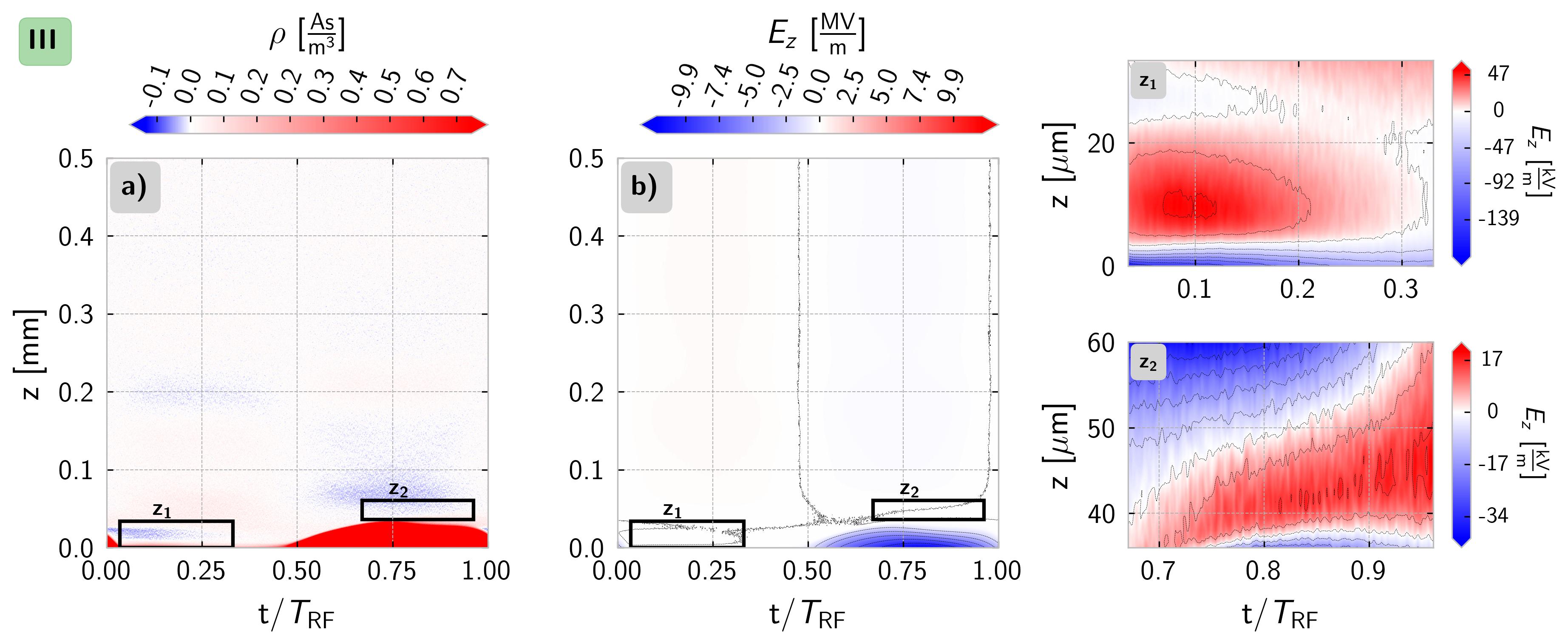}
        \caption{Spatially and temporally resolved charge density $\rho$ (a)) and electric field $E_z$ (b)) for a constricted plasma.
            The panels z$_1$ and z$_2$ show magnifications of the electric field in the correspondingly marked zones.
            The color bar of panel a) is capped at $0.7\, \mathrm{As/m^3}$ to make values of the negative space charges appear in the plot.
            The maximum value of the simulated space charge is at $4.4\, \mathrm{As/m^3}$.
            Simulated parameters: $f_\mathrm{RF} = 18\,$MHz, $V_\mathrm{RF} = 350\,$V, $p = 101325\,$Pa, $x_\mathrm{N_2} = 0.001$, $T_\mathrm{g} = 300\,$K.}
        \label{fig:constricted_charge_density}
    \end{figure}

\begin{figure}[t!]
        \centering
        \includegraphics[width=\textwidth]{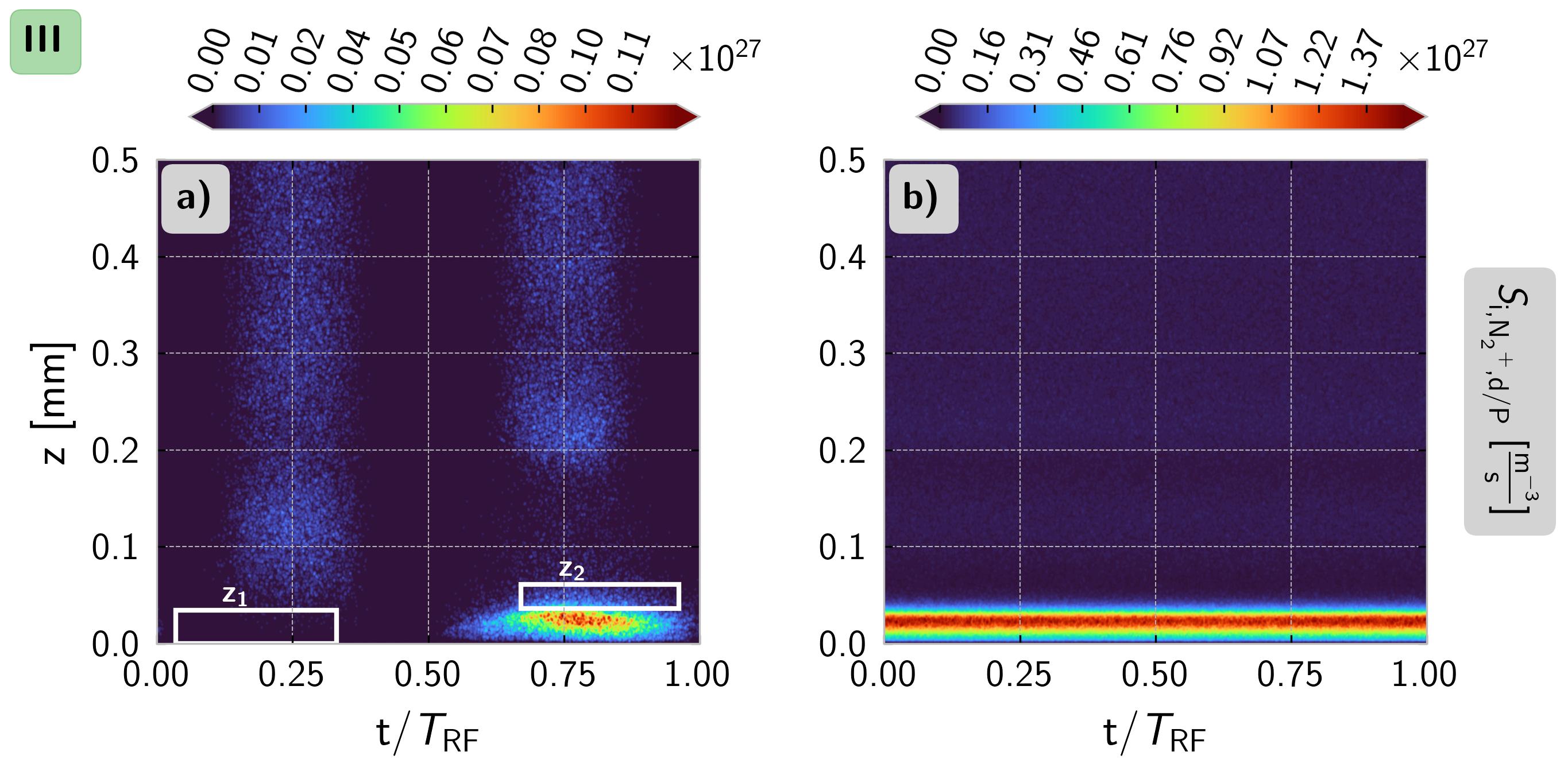}
        \caption{Ionization dynamics of a constricted plasma in spatial and temporal resolution.
            Nitrogen ionization is split into two components: a) electron impact ionization and b) Penning ionization.
            Panel a) includes the previously discussed zones z$_1$ and z$_2$ for reference.
            Simulated parameters: $f_\mathrm{RF} = 18\,$MHz, $V_\mathrm{RF} = 350\,$V, $p = 101325\,$Pa, $x_\mathrm{N_2} = 0.001$, $T_\mathrm{g} = 300\,$K.}
        \label{fig:constricted_ionization}
    \end{figure}

    For the constricted mode, the overshooting of the electrons can only be seen in terms of the charge density.
    (At plasma densities in the order of $10^{19}\, \mathrm{m^{-3}}$ tiny disparities between the densities of charged particles already cause relevant amounts of space charge.)
    Said charge density is presented in spatial and temporal resolution in figure \ref{fig:constricted_charge_density} a).
    At first glance, one will notice that the most significant accumulation of charges is the positive space charge during the expanded sheath phase ($t/T_\mathrm{RF} \approx 0.5 - 1.0$).
    The plot highlights two zones (z$_1$ and z$_2$) to focus on the dynamics connected to the overshooting of the electrons.
    The first inside zone one is in the same place as before, namely during sheath collapse.
    The second one inside zone two is located in front of the expanded sheath and is unique to constricted discharges.
    As discussed in section \ref{general}, a unique feature of constricted discharges is their plasma density profile.
    This plasma density profile exhibits maximum densities directly in front of the electrodes.
    Hence, there is a density gradient both towards the electrode and towards the plasma bulk (c.f., fig. \ref{fig:ave_densities} c)).
    Accordingly, electrons are now able to overshoot the ion density when moving in both directions:
    \begin{enumerate}
        \item In zone one, the sheath is collapsed, and electrons are pulled towards the electrode.
            Due to the high number of electrons that need to reach the electrode, a field reversal is necessary to drive the corresponding current.
            The negative charge density forms as a consequence.
        \item In zone two, the sheath is expanded, and electrons are pushed from the electrode back to the bulk.
            Additionally, secondary electrons are generated and, as we will discuss in figure \ref{fig:constricted_ionization}, electron and ion density grow rapidly by the extreme amount of ionization.
            Electrons now overshoot the ion density again because the same current that was driven towards the electrode in zone one now flows towards the bulk.
            Another field reversal occurs right before the expanding sheath that tries to pull the electrons back toward the maximum ion density.
    \end{enumerate}
    
    This makes two zones of negative space charge per half-period for the constricted mode.
    The electric field, shown in figure \ref{fig:constricted_charge_density} b), and the magnifications on the right side of the figure support the above description of these dynamics.
    (Please note that panels b), z$_1$ and z$_2$ each have individual color bars and scales).
    Overall, the space charge and electric field structure resemble the characteristic structures found in dc discharges \cite{raizer_gas_1997, bogaerts_hybrid_1995, fiala_two-dimensional_1994}.
    The most notable difference between these dc discharges is that the polarity switches each half-period due to the sinusoidal excitation, causing the electrodes to change their roles accordingly.
		
    \begin{figure}[t!]
        \centering
        \includegraphics[width=\textwidth]{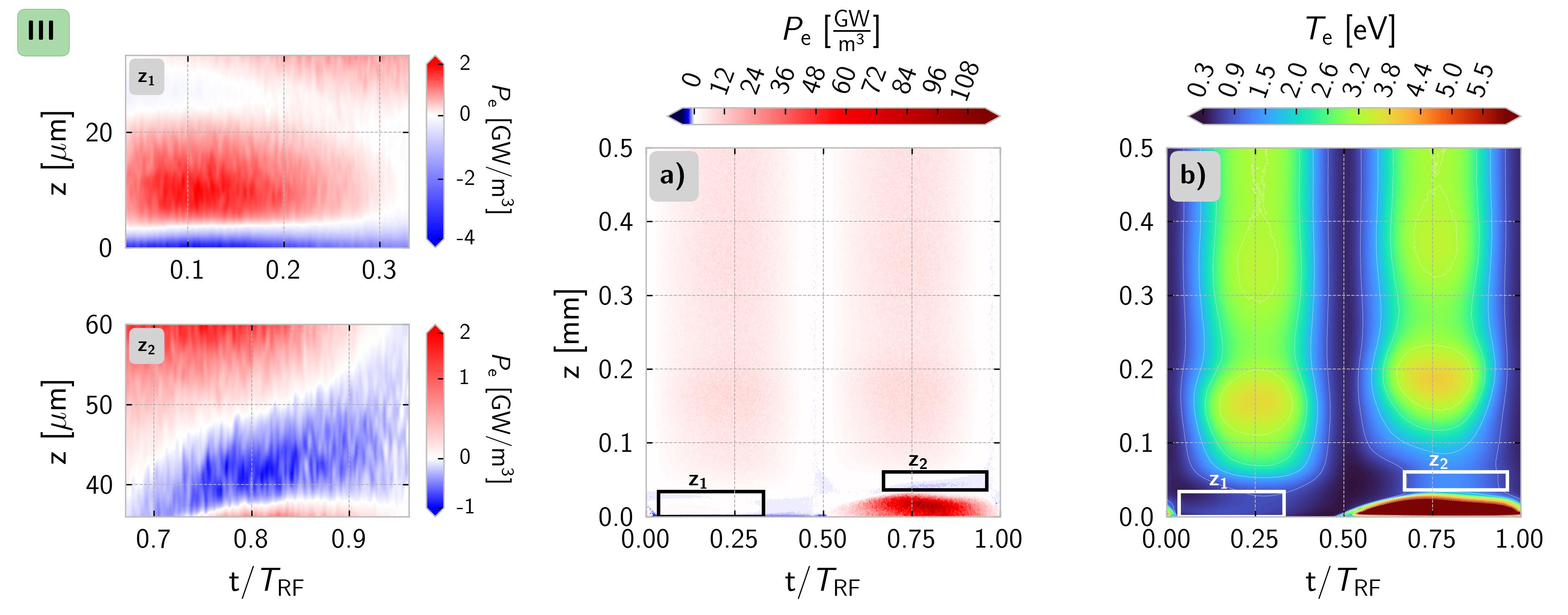}
        \caption{Spatially and temporally resolved electron power density $P_\mathrm{e}$ (a)) and electron temperature $T_\mathrm{e}$ (b)) for a constricted plasma.
            The panels z$_1$ and z$_2$ show magnifications of the electron power density in the correspondingly marked zones.
            Simulated parameters: $f_\mathrm{RF} = 18\,$MHz, $V_\mathrm{RF} = 350\,$V, $p = 101325\,$Pa, $x_\mathrm{N_2} = 0.001$, $T_\mathrm{g} = 300\,$K.}
        \label{fig:constricted_power_dynamics}
    \end{figure}

    The power dynamics shown in figure \ref{fig:constricted_power_dynamics}, especially the absorbed power by the electrons $P_\mathrm{e}$ in panel a), stress the statement that constricted plasmas centralize their dynamics inside the sheath region.
    Panel a) shows that the leading share of the energy is dissipated inside the sheath region.
    A significantly smaller portion of the absorbed power is directed into the bulk region.
    The dynamics associated with the two field reversals described earlier are nearly invisible.
    Thus, the panels z$_1$ and z$_2$ provide magnified versions of the power dynamics in the regions of these field reversals.
    In zone one, electrons are accelerated towards the electrode and gain energy during the sheath reversal in the collapsing sheath phase.
    This dynamic is similar to the other discharge modes discussed in section \ref{Dynamics1}.
    In zone two, electrons are accelerated back towards the expanding sheath.
    This acceleration opposes the current's direction, which currently flows toward the opposing electrode.
    The interaction with this second field reversal leads to electrons losing energy.

    The electron temperature $T_\mathrm{e}$ for this case, shown in panel b), also stresses these dynamics.
    In zone one, a tiny local electron density maximum is visible directly in front of the electrode.
    Zone two looks similar to the Faraday dark space of a dc discharge \cite{raizer_gas_1997, fiala_two-dimensional_1994}.
    Below ($z < 0.050\,$mm), electrons gain a vast amount of energy inside the sheath.
    This sheath can be understood as the momentary cathode layer (note: a constricted plasma here still is an RF-driven plasma and switches polarity).
    Above zone two ($z > 0.15\,$mm), electrons are contained inside the plasma bulk and gain a small amount of energy again.
    This energy gain is reflected in an increase in the electron temperature, and the plasma bulk of a constricted plasma can be understood as a positive column.
    The bulk region exists, yet mostly to electrically close the remaining discharge gap \cite{gudmundsson_foundations_2017}.
		
    Figure \ref{fig:constricted_ionization} shows the nitrogen ionization dynamics of the constricted mode in spatiotemporal resolution.
    The ionization rate is separated into two contributions, as in the previous section.
    Electron impact ionization of nitrogen (panel a)), and Penning ionization on nitrogen (panel b)).
    For reference, zones one and two are marked by white boxes in panel a).
    Overall, the main share of the ionization again occurs during the sheath expansion and inside the boundary sheath.
    These dynamics support the interpretation of constricted plasmas as atmospheric pressure versions of the $\gamma$-mode \cite{dunnbier_stability_2015, farouk_atmospheric_2008, laimer_investigation_2005, laimer_glow_2006, yang_comparison_2005, zhang_excitation_2009}.
    This dc-like discharge structure manifests in the ionization profiles.
    We look at panel a) at $t/T_\mathrm{RF} \approx 0.75$ as a reference.
    The lower electrode is the momentary cathode of the discharge.
    Similar to the ionization patterns in a dc-discharge \cite{raizer_gas_1997, bogaerts_hybrid_1995, fiala_two-dimensional_1994, gudmundsson_foundations_2017} there is ionization within the boundary sheath.
    This ionization maximum is comparable to the ionization inside a cathode layer.
    Next, no ionization ($z \approx 0.1 - 0.2\,$mm) leads to a dark space.
    This dark space results from the energy loss of electrons inside the field reversal of zone two.
    Beyond this dark space ($z > 0.2\,$mm) lies a zone of weak ionization.
    As shown before, this position is inside the plasma bulk, and the region is comparable to the positive column of an actual dc-discharge \cite{raizer_gas_1997, bogaerts_hybrid_1995, fiala_two-dimensional_1994, gudmundsson_foundations_2017}.
    Due to the depiction, we switch to $t/T_\mathrm{RF} \approx 0.25$ to continue the description and reach the momentary anode.
    At this point, the ionization within the plasma bulk ends at $z \approx 0.05\,$mm.
    Approaching the electrode that functions as a momentary anode, there is a small zone without notable ionization.
    This region, mainly located inside zone one, is comparable to the anode dark space \cite{gudmundsson_foundations_2017}.
    A structure comparable to an anode glow does not show in the simulations.
    Yet, this might be connected to insufficient resolution in terms of superparticles (i.e., a drastically increased amount of superparticles might lead to a weak local maximum due to the field reversal of zone one).

    The Penning ionization, shown in panel b), still contains no temporal structures and is entirely diffuse.
    However, it is connected to the trend discussed in the previous section.
    For the transition from non-neutral discharges to quasi-neutral plasmas, we discussed that the profile of the Penning ionization gradually transforms from a diffusion-like central peak profile into a bimodal one (c.f., bottom row of fig. \ref{fig:ion_dynamics}).
    The higher the plasma density, the more separated the peaks of this bimodal profile become.
    For constricted plasmas, the profile of the Penning ionization reaches the extreme case.
    The main contribution is constricted to the minuscule sheath regions.

%%%%%%%%%%%%%%%%%%%%%%
%%%%%%%%%%%%%%%%%%%%%%
%%%%%%%%%%%%%%%%%%%%%%
\section{Conclusion} \label{conclusion}
    This paper presents that the COST-Jet can be operated in three different discharge modes.
    There is the non-neutral mode at low input power.
    The defining characteristic of these discharges is the absence of a quasi-neutral plasma bulk.
    Electrons are strongly modulated in space and time, and the whole electron dynamics centralizes around the movement of the electron group.
    Next, the discharge enters the quasi-neutral mode at moderate power, and the, for capacitively coupled plasmas (CCPs) classical, bulk-sheath structure develops.
    For quasi-neutral plasmas, the electron dynamics are comparable to low-pressure CCPs, and two heating modes (i.e., the $\Omega$- and the Penning mode) can be differentiated.
    If the applied voltage suffices, the plasma jumps to a third discharge mode, the constricted mode.
    The constricted mode is a high-power and high-density mode with extreme and extremely localized dynamics.
    Moreover, although a radio frequency drives the constricted plasma, it exhibits dynamics more similar to the inherent dynamics of direct-current (dc) discharges.
    These dc-like dynamics further distinguish the constricted mode from the previous two.
    
    It was shown that the appearance of the discharge modes is parameter-dependent by applying both a frequency and a voltage variation.
    Varying, for example, the frequency causes the discharge to transit between the non-neutral and the quasi-neutral mode.
    Like a dc discharge, the constricted mode requires a minimal applied voltage and does not appear while varying the frequency in our work.
    At the same time, a continuous mode transition from I to II to III is not guaranteed.
    A non-neutral discharge is not bound to transit to the quasi-neutral mode.
    During the voltage variation presented for the $12\,$MHz cases, for example, the discharge stays non-neutral until it directly jumps to the constricted mode.
    Based on the simulated data, this behavior is expected for frequencies below $18\,$MHz.

    Despite their individual dynamics, this work presents that negative space charges and coinciding field reversals significantly contribute to the electron dynamics of all discharge modes.
    In general, we saw electrons overshoot the total ion density whenever there were steep gradients in the latter.
    This overshoot occurs for all discharge modes during the collapsing sheath phase, creating a field reversal.
    Electrons are accelerated towards the electrode, gain noticeable energy, and cause the field reversal structure in the ionization profile.
    Due to the unique density profile in constricted plasmas, this process repeats during the expanding sheath phase.
    However, the field reversal then opposes the current's direction, and the mechanism effectively leads to a dark-space-like region.

    Experimental results support the findings presenting the constricted mode as a distinct operation mode.
    We show that the constricted mode can expand along almost the entire electrode area.
    Additionally, phase-resolved optical emission spectroscopy qualitatively validated the simulated data.
    However, we must stress that the COST-Jet design is not well suited for the constricted mode.
    A modified version of the COST-Jet was designed to allow for electrode cooling.
    Solely when cooling is applied, a COST-Jet-like device can operate in the constricted mode described in this paper.

%%%%%%%%%%%%%%%%%%%%%%
%%%%%%%%%%%%%%%%%%%%%%
\ack  Funding by the German Research Foundation DFG in the frame of SFB 1316 (project number 327886311, projects A8, A4, and A5) is gratefully acknowledged.

\section*{\scriptsize{ORCID IDs}}
\begin{multicols}{2}
    \scriptsize{\noindent M. Klich: \href{https://orcid.org/0000-0002-3913-1783}{https://orcid.org/0000-0002-3913-1783}\\}
    \scriptsize{\noindent D. Schulenberg: \href{https://orcid.org/0000-0002-4086-8678}{https://orcid.org/0000-0002-4086-8678}\\}
    \scriptsize{\noindent S. Wilczek: \href{https://orcid.org/0000-0003-0583-4613}{https://orcid.org/0000-0003-0583-4613}\\}
    \scriptsize{\noindent M. Vass: \href{https://orcid.org/0000-0001-9865-4982}{https://orcid.org/0000-0003-2384-1243}\\}
    \scriptsize{\noindent T. Bolles: \href{https://orcid.org/0009-0007-7374-7286}{https://orcid.org/0009-0007-7374-7286}\\}
    \scriptsize{\noindent I. Korolov: \href{https://orcid.org/0000-0003-2384-1243}{https://orcid.org/0000-0003-2384-1243}\\}
    \scriptsize{\noindent J. Schulze: \href{https://orcid.org/0000-0001-7929-5734}{https://orcid.org/0000-0001-7929-5734}\\}
    \scriptsize{\noindent T. Mussenbrock: \href{https://orcid.org/0000-0001-6445-4990}{https://orcid.org/0000-0001-6445-4990}}\\
    \scriptsize{\noindent R. P. Brinkmann: \href{https://orcid.org/0000-0002-2581-9894}{https://orcid.org/0000-0002-2581-9894}}
\end{multicols}

\printbibliography

\end{document}